\documentclass[aps,pra,twocolumn,superscriptaddress,10pt,nofootinbib]{revtex4-2}
\usepackage{multirow,amsthm,amssymb,amsbsy,amsmath,epsfig,epstopdf,bm,float}
\DeclareMathOperator{\Tr}{Tr}
\usepackage{enumitem}
\usepackage{graphicx}
\usepackage{booktabs}
\usepackage{amsmath}
\usepackage{verbatim}
\usepackage{dcolumn}
\usepackage{color}
\usepackage[dvipsnames]{xcolor}
\usepackage{mathtools}
\usepackage[normalem]{ulem}
\usepackage{soul}
\usepackage{hyperref}
\usepackage{braket}
\usepackage[export]{adjustbox}
\DeclarePairedDelimiter\abs{\lvert}{\rvert}%

\newcommand\restr[2]{{
  \left.\kern-\nulldelimiterspace 
  #1 
  \vphantom{\big|} 
  \right|_{#2} 
  }}

\newcommand{\mar}[1]{\textcolor{black}{#1}}

\begin{document}

\title{Symmetric dilations of \mar{multi-qubit} Pauli channels and semigroups
}

\author{Marco Cattaneo}
\email{marco.cattaneo@uab.cat}
\affiliation{Department of Physics, University of Helsinki, P.O. Box 43, FI-00014 Helsinki, Finland}
\affiliation{GIQ group, Department of Physics, Autonomous University of Barcelona, E-08193 Bellaterra, Spain}

\date{\today}

\begin{abstract}
We explore the symmetry properties of Stinespring dilations of Pauli channels, addressing both the generic $n$-qubit case and specific examples such as the single-qubit phase damping and depolarizing channels. For each scenario, we derive the representation of the Pauli group acting on the Hilbert space of the environment. We then focus on dilations that are continuous in time and driven by a Hamiltonian, and on collision models that generate a Pauli dynamical semigroup in the limit of fast collisions. First, we complement some recent general results on these types of dilations (\href{https://journals.aps.org/pra/abstract/10.1103/PhysRevA.111.022209}{M. Cattaneo, Phys. Rev. A 111, 022209 (2025)}) with some additions and clarifications, including the case of covariant channels with strongly conserved quantities. Next, we show that the covariance property of Pauli channels imposes strong constraints on both the dilation Hamiltonian and the initial state of the environment, and demonstrate how these constraints can be exploited to explicitly construct the time-dependent dilations in all considered cases. In particular, for $n$-qubit Pauli channels we fully characterize the symmetric dilation Hamiltonians and show that, with time-dependent couplings, they generate any time-dependent Pauli channel. Our results are relevant for the quantum simulation of multi-qubit Pauli channels via unitary dilations and of multi-qubit Pauli semigroups via collision models, both in the laboratory and on quantum computers.
\end{abstract}
\maketitle

\section{Introduction}\label{sec:intro}
The celebrated Stinespring dilation theorem \cite{Stinespring1955,Paulsen2003} is one of the cornerstones of quantum information theory, as it allows us to interpret any physical quantum evolution as a unitary dynamics on an extended Hilbert space. By ``physical evolution'' we mean a generic completely positive trace-preserving (CPTP) linear map, which we refer to as a \textit{quantum channel} \cite{nielsenchuang}. Moreover, the extended Hilbert space of the dilation can be split into the Hilbert space of the \textit{system} on which the original quantum channel acts and the additional Hilbert space of the \textit{environment}.

The most immediate application of the Stinespring dilation theorem is the design of simple strategies for the quantum simulation of open systems and quantum channels. Indeed, given a channel we aim to simulate, we can always find a suitable Stinespring dilation thereof that can be experimentally implemented through controlled unitary evolutions in a laboratory or through a sequence of quantum gates on a quantum computer. This protocol has already been demonstrated on different experimental platforms \cite{Barreiro2011,Schindler2013,Xin2017,Han2021} and digital quantum computers \cite{Wei2018,DelRe2020,Garcia-Perez2020,David2024}.

A particularly interesting case is the quantum simulation of open quantum systems, and specifically of Markovian dynamics driven by quantum dynamical semigroups \cite{rivas2012open,vacchini2024open}. In general terms, it is impossible to simulate a quantum dynamical semigroup by means of a Stinespring dilation with a finite-dimensional Hilbert space of the environment \cite{VomEnde2023}. However, dynamical semigroups can be approximately simulated by means of \textit{quantum collision models} \cite{Campbell2021a,Ciccarello2021,Cattaneo2022d,Cusumano2022,Lacroix2024a}, which consist of repeated interactions between the open system of interest and different ancillary particles (typically qubits) of the environment. In the limit of fast collisions, it can be shown that collision models can efficiently simulate any quantum dynamical semigroup, including collective dynamics \cite{Cattaneo2021}. These protocols have also been tested on near-term quantum computers \cite{Burger2022,Erbanni2023,Cattaneo2023}.

\textit{Covariant} quantum channels are of great importance in quantum information \cite{Vacchini2009,Memarzadeh2022} and open quantum systems \cite{Holevo1993,Holevo1996,Buca2012,Albert2014}, where they are also known as \textit{weakly symmetric} open dynamics. In this paper, we focus on \mar{$n$}-qubit \textit{Pauli channels} \cite{Bengtsson2017} that can be shown to be covariant with respect to the Pauli group \cite{nielsenchuang,Poshtvan2022}. Pauli channels are a particularly relevant subclass of quantum channels, as many of the most fundamental noise maps, such as phase damping and depolarizing channels \cite{nielsenchuang}, are Pauli channels. Moreover, Pauli channels play a key role in quantum computing, as a well-established  technique known as ``Pauli twirling'' or ``randomized compiling'' can transform, under certain assumptions, any noise channel into a Pauli channel \cite{Wallman2016}. Understanding the properties of these channels is then crucial for efficient noise estimation on quantum computers \cite{Flammia2020,Chen2025}. Pauli channels also play a key role in the theory of quantum error correction \cite{gutierrez2015comparison,Wagner2022}, and techniques for their digital quantum simulation have been recently proposed \cite{Basile2024Pauli}.

In this work, we study the physical dilations of Pauli channels from the perspective of their covariance with respect to the Pauli group. This point of view is particularly interesting because of a theorem that guarantees the existence of a covariant Stinespring dilation with a proper group representation on the Hilbert space of the environment for any covariant quantum channel \cite{Scutaru1979,Keyl1999}. This theorem has been extended to fully symmetric physical dilations\footnote{In this work we refer to ``Stinespring dilations'' as dilations defined by the Hilbert space of the environment and an isometry operator, while ``physical dilations'' further specify a physical unitary operator and an initial state of the environment. We refer to Sec.~\ref{sec:dilations} for further clarifications.} in later works \cite{MarvianMashhad2012,Faist2021}.  

Given minimal dilations of the single-qubit phase damping, depolarizing, and generic \mar{multi}-qubit Pauli channels, in this paper we explicitly obtain the corresponding representations of the Pauli group on the space of the environment. Next, we assume that the channels depend continuously on time, and their dilation can be implemented through a standard unitary evolution driven by a Hamiltonian or, in the case of Pauli semigroups \cite{Puchaa2019,Jagadish2020,Chruscinski2024}, through fast-collision models. In these scenarios, we recently showed for general covariant channels that some strong constraints on the structure of the dilation Hamiltonian and on the initial state of the environment emerge \cite{Cattaneo2025Dilations}: \mar{for dilations whose environment representation is time independent}, the initial state of the environment \textit{must} be invariant with respect to such group representation, and the dilation Hamiltonian \textit{must} commute with the group representation on the subspace spanned by the dynamics, turning the existence theorem of Refs.~\cite{Scutaru1979,Keyl1999,MarvianMashhad2012,Faist2021} into a necessary condition.

Here, we clarify some aspects of \cite{Cattaneo2025Dilations} that were not fully addressed there, and then apply these results to Pauli channels.  We show how the constraints imposed by covariance with respect to the Pauli group (and $SU(2)$ in the case of the depolarizing channel) help us explicitly construct dilation Hamiltonians and characterize their symmetry properties. \mar{In particular, for $n$-qubit Pauli channels we obtain a complete characterization of the symmetric physical dilations: the environment representation depends only on which Pauli operators appear in the channel, the initial state of the environment is uniquely fixed, and every symmetric dilation consists of a free evolution of the environment, which can be seen as a continuous-time quantum walk \cite{Farhi1998}, followed by coherently controlled Pauli operations on the system. With time-dependent couplings, these dilations generate any time-dependent $n$-qubit Pauli channel, and the same construction provides collision models for any $n$-qubit Pauli semigroup.} These results are of immediate interest for the simulation of parametrized Pauli channels that are continuous in time \cite{UrRehman2024}, including unitary dilations thereof \cite{Acevedo2026} and Pauli dynamical semigroups. 

The paper is structured as follows. First, we  discuss the dilations of covariant channels in Sec.~\ref{sec:dilations}, where we also present some new results on dilations driven by time-independent Hamiltonians. Pauli channels are then introduced in Sec.~\ref{sec:Pauli}. \mar{For single-qubit channels we employ the standard language of Pauli matrices and Bloch vector representation, which has a direct physical interpretation; for multi-qubit channels we will instead adopt the more abstract binary symplectic representation of Pauli strings \cite{dion2024efficiently}.} Next, Sec.~\ref{sec:results} is devoted to the derivation of the symmetric dilations of \mar{single-qubit} Pauli channels and their consequences for time-continuous dilations.  \mar{These concrete examples illustrate the main ideas and their direct physical implications, while the general multi-qubit case is addressed in Sec.~\ref{sec:resultsMulti} at a more abstract level, which recovers the single-qubit results in the appropriate limit.} Finally, we draw some concluding remarks in Sec.~\ref{sec:conclusions}.

\section{Dilations of covariant maps}
\label{sec:dilations}

\subsection{Stinespring and physical dilations}

Given a quantum system with Hilbert space $\mathcal{H}_S$ and a quantum channel $\phi$ acting on it, we can always find a \textit{Stinespring dilation} of $\phi$, which is defined through a Hilbert space of the environment $\mathcal{H}_E$ and an isometry $V:\mathcal{H}_S\rightarrow \mathcal{H}_S \otimes \mathcal{H}_E$, with $V^\dagger V=\mathbb{I}_S$, such that
\begin{equation}
    \label{eqn:StinespringDilation}
    \phi[\rho_S]=\Tr_E[V \rho_S V^\dagger].
\end{equation}
A more rigorous formulation of the Stinespring theorem can be found in \cite{Stinespring1955,Paulsen2003}.

To connect this theorem to physical dilations that we may be able to implement in a lab, we introduce the dilation unitary $U:\mathcal{H}_S\otimes \mathcal{H}_E \rightarrow\mathcal{H}_S\otimes \mathcal{H}_E$ and the initial pure state of the environment $\ket{\psi_E}$, such that 
\begin{equation}\label{eqn:fromVtoU}
    V\ket{\varphi_S} = U\ket{\varphi_S}\otimes \ket{\psi_E}\quad \forall\,\ket{\varphi_S}\in\mathcal{H}_S.
\end{equation}
Then, a \textit{physical dilation} of $\phi$ assumes the more familiar form \cite{nielsenchuang,Caruso2006}
\begin{equation}\label{eqn:physicalDilation}
    \phi[\rho_S]=\Tr_E[U \rho_S\otimes\ket{\psi_E}\!\bra{\psi_E}U^\dagger]. 
\end{equation}

Throughout the paper we will work  with pure states of the environment only. Extensions of our results to mixed states of the environment are quite straightforward through a purification procedure \cite{nielsenchuang,Caruso2006,Cattaneo2025Dilations}.

Next, the notion of \textit{minimal Stinespring dilation} is quite important for our goals. A Stinespring dilation defined by $V$ and $\mathcal{H}_E$ is minimal if \cite{Paulsen2003}
\begin{equation}\label{eqn:minimalDil}
\begin{split}
    &Span\{a\otimes\mathbb{I}_E V \ket{\varphi_S}, \,\forall a\in\mathcal{B}(\mathcal{H}_S),\,\forall\ket{\varphi_S}\in\mathcal{H}_S\} \\
    &=\mathcal{H}_S\otimes\mathcal{H}_E,    
\end{split}
\end{equation}
where $\mathcal{B}(\mathcal{H}_S)$ is the space of bounded operators on $\mathcal{H}_S$. It can be shown that all minimal Stinespring dilations of a channel $\phi$ are unitarily equivalent \cite{Paulsen2003}.

A minimal dilation can also be characterized as follows. Suppose that the channel $\phi$ can be represented through the Kraus operators $\{K_j\}_{j=1}^{r_K}$, where $r_K$ is the \textit{Kraus rank} of $\phi$, i.e., the minimal number of Kraus operators in the representation of $\phi$, which is equal to the rank of the associated Choi matrix \cite{Watrous2018}. Then, a minimal Stinespring dilation of $\phi$ with environment $\mathcal{H}_E$ is characterized by \cite{Watrous2018,Lancien2024}
\begin{equation}\label{eqn:krausRank}
    dim(\mathcal{H}_E)=r_K.
\end{equation}

The construction of a minimal physical dilation is then quite straightforward. We introduce a basis $\{\ket{e_j}\}_{j=1}^{r_K}$ of $\mathcal{H}_E$, the initial state of the environment $\ket{\psi_E}$, and define the unitary operator $U$ such that \cite{nielsenchuang}
\begin{equation}
\label{eqn:explicitConstructionDilationKraus}
U\ket{\varphi_S}\otimes\ket{\psi_E} = \sum_{j=1}^{r_K} K_j \ket{\varphi_S}\otimes\ket{e_j}. 
\end{equation}
We immediately realize that $U$ and $\ket{\psi_E}$ give rise to a correct minimal  physical dilation of $\phi$ on $\mathcal{H}_S\otimes\mathcal{H}_E$, written as in \eqref{eqn:physicalDilation}. The same formula can be used to identify the minimal dilation isometry $V=U\ket{\psi_E}$ (with abuse of notation).

\subsection{Existence of a covariant dilation for any covariant channel}

We now focus on Stinespring dilations of channels that are \textit{covariant} with respect to a group $G$. Given a generic unitary representation $\pi_S(g)$ of $G$ on $\mathcal{H}_S$, a channel $\phi$ is covariant with respect to $G$ if, for any state $\rho_S$, 
\begin{equation}\label{eqn:covarianceProperty}
  \phi[\pi_S(g) \rho_S \pi_S^\dagger(g)]=\pi_S(g) \phi[\rho_S] \pi_S^\dagger(g)  \text{ for all }g\in G.
\end{equation}

A theorem due to Scutaru \cite{Scutaru1979}, which was later rediscussed by Keyl and Werner \cite{Keyl1999}, states that, \textit{for any covariant channel} $\phi$ with respect to a group $G$ with finite-dimensional unitary representation $\pi_S(g)$, \textit{there exists} a Stinespring dilation with isometry $V$ and Hilbert space of the environment $\mathcal{H}_E$, and a representation $\pi_E(g)$ of $G$ on $\mathcal{H}_E$, such that 
\begin{equation}\label{eqn:covariantDilationKeyl}
    V \pi_S(g) = \pi_S(g)\otimes \pi_E(g) V \quad \text{ for all } g\in G.
\end{equation}
More specifically, it can be shown that \textit{any} minimal Stinespring dilation with $(V,\mathcal{H}_E)$ of a covariant channel satisfies Eq.~\eqref{eqn:covariantDilationKeyl} \cite{Keyl1999}. 

A related theorem for physical dilations was later proven by Marvian \cite{MarvianMashhad2012}, who showed that, for any covariant channel $\phi$, \textit{there always exists} a physical dilation $U$, an initial state of the environment $\ket{\psi_E}$, and a representation $\pi_E(g)$ of $G$ on $\mathcal{H}_E$ such that, for all $g\in G$,
\begin{equation}
    \pi_E(g)\ket{\psi_E}=\ket{\psi_E}, \quad[U,\pi_S(g)\otimes\pi_E(g)]=0.
\end{equation}
In contrast to \cite{Scutaru1979,Keyl1999}, the required dimension of the Hilbert space of the environment  in the constructive proof of \cite{MarvianMashhad2012} is $r_K+1$, where $r_K$ is the Kraus rank. Therefore, the physical dilation of the theorem in \cite{MarvianMashhad2012} is not minimal by construction. A similar theorem was proven by Faist et al. for quantum channels that are covariant with respect to time evolution \cite{Faist2021}. 

\subsection{Symmetric dilations of time-dependent channels}
\label{sec:symmetricDilTime}
We have recently explored the consequences of the theorem expressed in \eqref{eqn:covariantDilationKeyl} for time-dependent channels with some specific types of physical dilations \cite{Cattaneo2025Dilations}. In particular, given a time-dependent channel $\phi(t)$, here we focus on two particularly interesting classes of physical dilations that reproduce $\phi(t)$ for all $t\geq 0$: 
\begin{itemize}
    \item Dilations driven by a time-independent Hamiltonian $H_I$, with unitary 
    \begin{equation} \label{eqn:unitaryTimeIndependentDil}
        U_I(t)=\exp(-i H_I t).
    \end{equation}
    Later on in the paper, we will also make some comments on time-dependent Hamiltonians.
    Note that throughout the paper we set $\hbar =1$. 
    \item Collision models in the regime of fast collisions, for which each collision can be modeled through the unitary 
    \begin{equation}\label{eqn:collisionUnitary}
        U_c(\Delta t)= \exp(-i H_c \Delta t), \quad\Delta t\ll 1 ,   \end{equation}
    for a Hamiltonian operator $H_c$ with dimensionless units. It can be shown that the repeated application of \eqref{eqn:collisionUnitary} on the system and a sequence of ancillary qubits, which are traced out after each collision, can generate any quantum dynamical semigroup on $\mathcal{H}_S$ in the proper limit $\Delta t\ll 1$. The interested readers can find more details, for instance, in \cite{Ciccarello2021,Campbell2021a,Cusumano2022,Cattaneo2022d,Cattaneo2021,Cattaneo2025Dilations}.
\end{itemize}

Moreover, we now restrict ourselves to minimal physical dilations with pure states of the environment. If the dilation is not minimal, then it can be reduced to a minimal dilation using \eqref{eqn:minimalDil} and all results will follow accordingly. For simplicity, we will now also focus on dilations driven by a time-independent Hamiltonian, as the case of collision models behaves essentially in the same way. 

We are interested in a channel $\phi(t)$ that can be dilated through a unitary driven by a time-independent Hamiltonian as in \eqref{eqn:unitaryTimeIndependentDil}:
\begin{equation}\label{eqn:physDilTimeDep}
    \phi(t)[\rho_S] = \Tr_E[U_I(t)\rho_S\otimes\ket{\psi_E}\!\bra{\psi_E}U_I^\dagger(t)]\quad\forall t\geq0,
\end{equation}
given an initial state of the environment $\ket{\psi_E}$. Moreover, we assume that the channel $\phi(t)$ is covariant with respect to the representation $\pi_S(g)$ of a group $G$ for all $t\geq 0$:
\begin{equation}\label{eqn:covarianceTimeDep}
    \phi(t)[\pi_S(g)\rho_S\pi_S^\dagger(g)]=\pi_S(g)\phi(t)[\rho_S]\pi_S^\dagger(g) \quad\forall g\in G.
\end{equation}

In the following, we consider dilations for which the isometry $V(t) = U_I(t)\ket{\psi_E}$ satisfies \eqref{eqn:covariantDilationKeyl} at all times with the same representation $\pi_E(g)$ of $G$ on $\mathcal{H}_E$, up to a possible trivial rotating phase (see the discussion below). For these dilations, it can be shown that \cite{Cattaneo2025Dilations}:
\begin{equation}
        \label{eqn:invariantState}
        \pi_E(g)\ket{\psi_E}=\ket{\psi_E}\quad\forall g\in G.
\end{equation}
In other words, the group representation $\pi_E(g)$ \textit{must} act trivially on the initial state of the environment $\ket{\psi_E}$. Moreover, 
\begin{equation}
    \label{eqn:symmetryDilationParallel}
     \restr{[\pi_S(g)\otimes\pi_E(g),H_I]}{\mathcal{K}_\parallel}=0 \quad \forall g\in G,
\end{equation}
 where $\mathcal{K}_\parallel$ is the subspace spanned by the dynamics:
 \begin{equation}\label{eqn:sigmaParallel}
\mathcal{K}_\parallel=Span\{U_I(t)\ket{\varphi_S}\otimes\ket{\psi_E},\;\forall\ket{\varphi_S}\in\mathcal{H}_S,\;\forall t\geq 0\}.
 \end{equation}
 Therefore, due to the covariance property of $\phi(t)$, \textit{any} such dilation of $\phi(t)$ driven by the unitary operator in \eqref{eqn:unitaryTimeIndependentDil} \textit{must} be symmetric with respect to the tensor product representation $\pi_S\otimes\pi_E$  of $G$ on the subspace spanned by the dynamics. If $G$ is a Lie group, Eq.~\eqref{eqn:symmetryDilationParallel} implies a set of conserved quantities of $U_I(t)$ on $\mathcal{K}_\parallel$.
 
 It can be shown that the subspace $\mathcal{K}_\parallel$ can be nicely characterized in terms of Krylov subspaces. In addition, all results stated above are valid also for fast-collision models, with the only difference that the subspace spanned by the dynamics is defined differently.  Finally, any dilation Hamiltonian $H_I$ that is symmetric only on the subspace $\mathcal{K}_\parallel$ can be properly modified to obtain a Hamiltonian $H'_I$ that generates the same physical dilation and is symmetric on the full Hilbert space $\mathcal{H}_S\otimes\mathcal{H}_E$. We refer the interested readers to \cite{Cattaneo2025Dilations} for all details and proofs. 

The way to find the representation $\pi_E(g)$ given $\phi$, $\pi_S(g)$, and a physical dilation $(U_I(t)$, $\ket{\psi_E})$ passes through the theorem in \eqref{eqn:covariantDilationKeyl} for minimal Stinespring dilations. We discuss this point in the following two subsections, which add some  considerations to the results in \cite{Cattaneo2025Dilations}.

 \subsubsection{Trivial rotating phase in the group representation on the environment}\label{sec:trivialRot}
The representation on the Hilbert space of the environment can be found by applying Eq.~\eqref{eqn:covariantDilationKeyl} to the covariance property \eqref{eqn:covarianceTimeDep}, after we define a corresponding time-dependent dilation isometry through (with abuse of notation)
\begin{equation}\label{eqn:dilationIsometryTimeDep}
    V(t) = U_I(t) \ket{\psi_E}.
\end{equation}

For such a minimal dilation, Eq.~\eqref{eqn:covariantDilationKeyl} guarantees the existence of a time-dependent $\tilde{\pi}_E(g,t)$ such that 
\begin{equation}
    \label{eqn:timeDependentKeyl}
V(t)\pi_S(g)=\pi_S(g)\otimes\tilde{\pi}_E(g,t) V(t)\quad\forall g \in G.
\end{equation}
An argument presented in \cite{Cattaneo2025Dilations} suggests that, for meaningful non-trivial dilations $U_I(t)$ that are driven by a time-independent Hamiltonian like in \eqref{eqn:unitaryTimeIndependentDil}, the representation $\pi_E$ is actually time-independent. This argument, however, neglects the possibility of the existence of a trivial rotating phase in $\tilde{\pi}_E(g,t)$, which we discuss here.  

Suppose that the dilation Hamiltonian can be decomposed as 
\begin{equation}
    H_I = H_I^{(\text{int})} + \mathbb{I}_S\otimes H_E,
\end{equation}
where $H_E$ is a free environment Hamiltonian that commutes with $H_I^{(\text{int})}$:
\begin{equation}
    [H_I^{(\text{int})},\mathbb{I}_S\otimes H_E] = 0. 
\end{equation}

The Hamiltonian $H_E$ has no role in the dilation, as is readily seen by using the cyclic property of the trace:
\begin{equation} 
\begin{split}
&\Tr_E[U_I(t)\rho_S\otimes\ket{\psi_E}\!\bra{\psi_E}U_I^\dagger(t)]\\ &=\Tr_E[e^{-i H_I^\text{(int)}t}\rho_S\otimes\ket{\psi_E}\!\bra{\psi_E}e^{i H_I^\text{(int)}t}].
\end{split}
\end{equation}
Then, $H_I^{(\text{int})}$ is the non-trivial part of the Hamiltonian, which we assume to yield a time-independent $\pi_E$ such that 
\begin{equation}
    \label{eqn:timeDependentKeylNonTrivial}
V_\text{int}(t)\pi_S(g)=\pi_S(g)\otimes\pi_E(g) V_\text{int}(t)\quad\forall g \in G,
\end{equation}
for the dilation isometry
\begin{equation}
    V_\text{int}(t)=e^{-i H_I^\text{(int)}t}\ket{\psi_E}.
\end{equation}
Then, it is easy to verify that the original relation  \eqref{eqn:timeDependentKeyl} for the full $H_I$ is still valid with a representation
\begin{equation}
    \tilde{\pi}_E(g,t) = e^{-i H_E t}\pi_E(g)e^{i H_E t},
\end{equation}
which has an additional rotating phase. 

From a physical perspective, the Hamiltonian $H_E$ is completely redundant, as it is useless for the sake of the dilation. However, it is important to recognize that the possibility of adding this trivial phase to the dilation Hamiltonian, and consequently to the environment representation $\tilde{\pi}_E$, exists. In this case, we notice that the key results in Eqs.~\eqref{eqn:invariantState} and~\eqref{eqn:symmetryDilationParallel} hold for $H_I^\text{(int)}$ and $\pi_E(g)$, but no longer for $H_I$ and $\tilde{\pi}_E(g,t)$. We will observe a simple example of this scenario for Pauli channels in Sec.~\ref{sec:results}.

Finally, we point out that this trivial rotating phase is different from the unitary freedom in the choice of $H_I$ and $\ket{\psi_E}$ for the construction of a unitary physical dilation driven by a time-independent Hamiltonian. It is indeed known that, given $H_I$ and $\ket{\psi_E}$, an analogous unitary physical dilation of $\phi(t)$ can be built through \cite{Acevedo2026}
\begin{equation}
    H_I'=\mathbb{I}_S\otimes W_E H_I \mathbb{I}_S\otimes W_E^\dagger,\quad \ket{\psi_E'}=W_E\ket{\psi_E},
\end{equation}
for a generic time-independent unitary $W_E$ on $\mathcal{H}_E$. The isometry associated with the new rotated dilation is
\begin{equation}
    V'(t)=\mathbb{I}_S\otimes W_E V(t).
\end{equation}
Then, if $\pi_E(g)$ is a time-independent representation of $G$ associated with $V(t)$, the standard relation \eqref{eqn:covariantDilationKeyl} holds for $V'(t)$ with the new time-independent representation
\begin{equation}
    \pi'_E(g)= W_E \pi_E(g) W_E^\dagger.
\end{equation}
Finally, Eqs.~\eqref{eqn:invariantState} and~\eqref{eqn:symmetryDilationParallel} hold for the rotated $\pi'_E(g)$, $H_I'$ and $\ket{\psi'_E}$.

 \subsubsection{Covariant channels with strongly conserved quantities}\label{sec:conservedQuantity}

To start with a simple example, let us consider a representation of $U(1)$ that can be written as 
\begin{equation}\label{eqn:U1rep}
    \pi_S(g)= \exp (i g J),\quad g\in\mathbb{R},
\end{equation}
for some observable $J$ on $\mathcal{H}_S$. It is well known that if a channel $\phi$ is covariant with respect to \eqref{eqn:U1rep},  then this does not imply that $J$ is a conserved quantity of the channel. For Markovian open quantum systems, this kind of symmetry is called ``weak'' \cite{Buca2012,Albert2014}. The same concept can be easily extended to any compact Lie group, see for instance \cite{Marvian2014}. 

Here we focus instead on ``strong symmetries'', for which $J$ commutes with all Kraus operators of the covariant channel $\phi$ \cite{Li2025},
\begin{equation}
    [J,K_j]=0,\quad j=1,\ldots, r_K.
\end{equation}
Then, $J$ is also a conserved quantity of the dynamics.
Using the above condition and Eq.~\eqref{eqn:explicitConstructionDilationKraus}, we observe 
\begin{equation}\label{eqn:isometryVJ}
    V J =J\otimes\mathbb{I}_E V,
\end{equation}
and as a consequence
\begin{equation}\label{eqn:strongCovarianceDilation}
    V \pi_S(g)= \pi_S(g)\otimes\mathbb{I}_E V,\quad \forall g\in U(1),
\end{equation}
where with an abuse of notation we interpret $g\in\mathbb{R}$ as an element of $U(1)$. Indeed, we notice that, if \eqref{eqn:isometryVJ} holds, then
\begin{equation}
    V J^2 = J\otimes\mathbb{I}_E V J = J^2\otimes\mathbb{I}_E V, 
\end{equation}
and analogously for any power of $J$, proving \eqref{eqn:strongCovarianceDilation}.

The argument above can be straightforwardly extended to any compact Lie group whose Lie algebra generators are strongly conserved, and to discrete groups such that $\pi_S(g)$ are not only unitary but also Hermitian operators, and in addition are strongly conserved quantities of the channel: 
\begin{equation}
[K_j,\pi_S(g)]=0, \quad j=1,\ldots,r_K,\;\forall g\in G.
\end{equation} In all these cases, by comparing Eq.~\eqref{eqn:strongCovarianceDilation} with~\eqref{eqn:covariantDilationKeyl}, we realize that the theorem on the existence of a covariant dilation has here a trivial consequence: for covariant channels with an associated strongly conserved quantity, the group representation $\pi_E$ in \eqref{eqn:covariantDilationKeyl} is just a direct sum of trivial representations, i.e., it maps any group element to the identity:
\begin{equation}\label{eqn:piEIdentityConserved}
    \pi_E(g)= \mathbb{I}_E \quad \forall g\in G.
\end{equation}
Therefore, in these cases (including strong symmetries in open systems \cite{Buca2012,Albert2014}) the covariance property of the map does not give us any relevant information and/or constraint on the symmetry properties of the dilations, apart from the immediate consequence that, for example, $J\otimes\mathbb{I}_E$ for $U(1)$-covariance must also be conserved at the level of the dilation (or analogously $\pi_S(g)\otimes\mathbb{I}_E$ for discrete symmetries where $\pi_S(g)$ is  a strongly conserved observable).

Finally, we observe that the same reasoning applies to the case where only some $\pi_S(g)$ are strongly conserved observables or, for Lie groups, where only some of the generators of the Lie algebra are strongly conserved. The corresponding $\pi_E(g)$ for these group elements will necessarily be the identity on $\mathcal{H}_E$. The phase damping channel analyzed in Sec.~\ref{sec:phaseDamping} is a clear example of this scenario. 

\section{Pauli channels} \label{sec:Pauli}

\subsection{Single-qubit channels}
 A quantum channel $\phi$ acting on a single-qubit space $\mathcal{H}_S$ is a \textit{Pauli channel} if it can be written as \cite{Bengtsson2017} 
\begin{equation}\label{eqn:PauliChannel}
    \phi[\rho_S] = \sum_{\alpha=I,x,y,z} p_\alpha \sigma_\alpha \rho_S \sigma_\alpha,\quad p_\alpha\geq 0, \quad\sum_\alpha p_\alpha = 1, 
\end{equation}
where $\sigma_i$ with $i=x,y,z$ are the Pauli matrices, and $\sigma_I = \mathbb{I}_2$. $\rho_S$ is a generic single-qubit density matrix. From \eqref{eqn:PauliChannel} we immediately observe that Pauli channels are unital, $\phi[\mathbb{I}_2]=\mathbb{I}_2$.
In addition, the Kraus operators of a Pauli channel are  given by 
\begin{equation}
    \label{eqn:KrausPauliChannel}
K_\alpha= \sqrt{p_\alpha} \sigma_\alpha.
\end{equation}
As a trivial consequence, Pauli channels are self-adjoint:
\begin{equation}
    \phi^\dagger = \phi.
\end{equation}
We point out that throughout the paper we employ the Greek indexes $\alpha=I,x,y,z$ when the identity matrix is included, and the Roman indexes $i=x,y,z$ when it is not. 

If we represent a qubit state $\rho_S$ on the Bloch sphere \cite{nielsenchuang} with $\vec{\sigma}=(\sigma_x,\sigma_y,\sigma_z)^T$, 
\begin{equation} \label{eqn:blochSphere}
    \rho_S = \frac{1}{2}\left(\mathbb{I}_2+\vec{r}\cdot \vec{\sigma}\right),
\end{equation}
a Pauli channel acts on this state through an independent scaling of each component of the Bloch vector $\vec{r}$. Specifically (see Appendix~\ref{appendix:propertiesPauli}), 
\begin{equation}\label{eqn:PaulionBloch}
    \phi[\rho_S] = \frac{1}{2}\left(\mathbb{I}_2+\vec{r'}\cdot \vec{\sigma}\right), 
\end{equation}
with
\begin{equation}\label{eqn:PauliVectorRescaling}
    r_i' = \lambda_i r_i
\end{equation}
and
\begin{equation}\label{eqn:PauliScalingComponents}
\begin{split}
    &\lambda_x = p_I +p_x -p_y-p_z, \\
    &\lambda_y = p_I -p_x +p_y-p_z,\\
    &\lambda_z = p_I -p_x -p_y+p_z.
\end{split}
\end{equation}

Analogously, a \textit{Pauli dynamical semigroup} $\phi(t)=\exp(\mathcal{L}t)$ is characterized by a Liouvillian superoperator \cite{rivas2012open,vacchini2024open} that can be expressed as \cite{Puchaa2019,Jagadish2020,Chruscinski2024} 
\begin{equation}\label{eqn:PauliSemigroup}
    \mathcal{L}[\rho_S]=\sum_{i=x,y,z}\gamma_i (\sigma_i \rho_S \sigma_i -\rho_S), \quad \gamma_i\geq 0.
\end{equation}
The semigroup $\phi(t)$ is a Pauli channel at any time $t$, as shown in Appendix~\ref{appendix:propertiesPauli}.

The \textit{Pauli group} $G_P$ for a single qubit is a finite  group of order 16 formed by the elements \cite{nielsenchuang}
\begin{equation}\label{eqn:PauliGroup}
    G_P = \{ \pm \sigma_{\alpha},\pm i\sigma_{\alpha}: \;\alpha=I,x,y,z\} .
\end{equation}
Any Pauli channel $\phi$ is \textit{covariant} with respect to the action of the Pauli group, meaning that it satisfies \eqref{eqn:covarianceProperty} where $\pi_S(g)$ is simply the ``defining'' representation of the Pauli group, i.e., 
\begin{equation}\label{eqn:DefiningPauli}
    \pi_S(g)= g \text{ for all }g\in G_P.
\end{equation}
For a Pauli semigroup driven by the Liouvillian $\mathcal{L}$, we can replace $\phi$ with $\mathcal{L}$ in \eqref{eqn:covarianceProperty}.
We refer the readers to Appendix~\ref{appendix:propertiesPauli} for a simple proof of the covariance property of Pauli channels. 

\mar{
\subsection{Multi-qubit channels}
}
\mar{
Given a Hilbert space $\mathcal{H}_S$ of $n$ qubits, we adopt the binary symplectic representation of Pauli strings \cite{dion2024efficiently}: we define an $n$-qubit  \textit{Pauli string} characterized by the tuple $a=(a^x|a^z)$, where $a^x,a^z\in\{0,1\}^n$ are binary strings of length $n$,  as an operator
\begin{equation}\label{nqubitPauliString}
    P_a = i^{a^x\cdot a^z} X^{a^x}Z^{a^z},
\end{equation}
where 
\begin{equation}
\begin{split}
    &X^s = \sigma_x^{s_1}\otimes\ldots\otimes \sigma_x^{s_n},\\
    &Z^s = \sigma_z^{s_1}\otimes\ldots\otimes \sigma_z^{s_n},
\end{split}
\end{equation}
given a generic $n$-dimensional string $s$, and $a^x\cdot a^z=\sum_{j=1}^n a^x_j a^z_j$. 
An $n$-qubit Pauli channel is then defined as
\begin{equation}\label{eqn:nQubitPauliChannel}
    \phi[\rho_S] = \sum_{a\in A} p_a P_a \rho_S P_a,\quad \sum_{a\in A} p_a =1, 
\end{equation}
where $A=\{a:\, p_a > 0\}$ is the support of the channel. The single-qubit case is recovered through $P_{(0|0)}=\mathbb{I}_2$, $P_{(1|0)}=\sigma_x$, $P_{(1|1)}=\sigma_y$, $P_{(0|1)}=\sigma_z$.}

\mar{
Next, we introduce the expression
\begin{equation}\label{eqn:omegaAB}
    \omega(a,b)= a^x\cdot b^z + a^z\cdot b^x = \sum_{j=1}^{n}\left(a^x_j\, b^z_j + a^z_j\, b^x_j\right).
\end{equation}
It is then easy to verify that
\begin{equation}\label{eqn:commutPaPb}
    P_a P_b = (-1)^{\omega(a,b)} P_b P_a.
\end{equation}
}

\mar{Similarly, we define an $n$-qubit Pauli dynamical semigroup $\phi(t)=e^{\mathcal{L}t}$ through the Liouvillian
\begin{equation}\label{eqn:nQubitPauliSemigroup}
    \mathcal{L}[\rho_S] = \sum_{a\neq 0} \gamma_a (P_a \rho_S P_a- \rho_S),\quad \gamma_a\geq 0,
\end{equation}
so that $\phi(t)$ is an $n$-qubit Pauli channel for all $t\geq 0$ (see Appendix~\ref{appendix:propertiesPauli}).  Note that, with abuse of notation, throughout the paper we are using $a=0$ to represent the binary string $(00\ldots 0|00\ldots 0)$. Any $n$-qubit Pauli channel, as well as the Liouvillian of any Pauli semigroup, is covariant with respect to the defining representation of the $n$-qubit Pauli group
\begin{equation}\label{eqn:nqubitPauliGroup}
    G_P^{(n)} = \{\pm P_a,\pm i P_a:\ a=(a^x|a^z),\ a^x,a^z\in\{0,1\}^n\}.
\end{equation}
}
    
\section{Dilations of single-qubit Pauli channels and their symmetries}\label{sec:results}
In this section we first analyze the specific examples of \mar{single-qubit} phase damping and depolarizing channels, and then study generic \mar{single-qubit} Pauli channels. For each of these cases, we first construct  minimal physical dilations from the Kraus representation of the channel and Eq.~\eqref{eqn:explicitConstructionDilationKraus}. Then, we assume that the channel is time-dependent and the dilation is driven by a time-independent Hamiltonian as in \eqref{eqn:unitaryTimeIndependentDil}. We explore the constraints on the structure of the dilation Hamiltonian induced by the Pauli covariance property of the channel, and construct explicit symmetric dilations. Finally, we also apply these results to the Hamiltonian of collision models for the simulation of the corresponding Pauli semigroup, according to \eqref{eqn:collisionUnitary}.

In this section we will present many results that can be verified through some algebraic manipulations. We will skip the tedious algebraic steps here, and refer the readers to a detailed \textsc{Mathematica} notebook for all proofs \cite{notebook}.

\subsection{Phase damping channel}
\label{sec:phaseDamping}
The phase damping channel with probability $p$ is defined as 
\begin{equation}\label{eqn:phaseDampDef}
    \phi_p^{\text{PD}}[\rho_S] = (1-p) \rho_S + p\sigma_z\rho_S\sigma_z,\quad 0\leq p\leq 1.
\end{equation}
Its Kraus operators are $K_0 = \sqrt{1-p}\,\mathbb{I}_2$, $K_1= \sqrt{p}\,\sigma_z$. 

A fundamental property of the phase damping channel is the (strong) conservation of $\sigma_z$:
\begin{equation}\label{eqn:sigmaZconserved}
    \phi_p^{\text{PD}}[\sigma_z]=\sigma_z.
\end{equation}
If we consider a single qubit with Hamiltonian $\frac{\Omega}{2} \sigma_z$, the phase damping channel describes dephasing in the absence of dissipation \cite{nielsenchuang}. 

\subsubsection{Minimal dilation from Kraus decomposition} \label{sec:minimalDilPhasDamp}
First, we find a minimal Stinespring dilation of the phase damping channel using Eq.~\eqref{eqn:explicitConstructionDilationKraus}. As this channel has two Kraus operators only, we need a two-dimensional environment Hilbert space, i.e., a qubit. As a basis of $\mathcal{H}_E$, we choose $\{\ket{1},\ket{0}\}$, with 
\begin{equation}
    \sigma_z\ket{1}=\ket{1},\quad \sigma_z\ket{0}=-\ket{0}.
\end{equation}
We use the same canonical basis for the system qubit. Then, we find the dilation isometry
\begin{equation}\label{eqn:VphaseDamp}
    V = \begin{pmatrix}
        \sqrt{1-p} & 0 \\
        \sqrt{p} & 0 \\
        0 & \sqrt{1-p}\\
        0 & -\sqrt{p}
    \end{pmatrix}.
\end{equation}
Note that in the codomain $\mathcal{H}_S\otimes\mathcal{H}_E$ we are using the canonical tensor product basis $\{\ket{11},\ket{10},\ket{01},\ket{00}\}$. We will use this convention throughout the paper.

Next, we look for the environment representation of the Pauli group $\pi_E$ through the relation \eqref{eqn:covariantDilationKeyl} with $V$ from \eqref{eqn:VphaseDamp}. We find 
\begin{equation}\label{eqn:repPhaseDamp}
\begin{aligned}
    &{\pi}_E(\pm\sigma_\alpha)={\pi}_E( \pm i\sigma_\alpha)= \mathbb{I}_2,\quad &\text{if }\alpha=I,z,\\ &{\pi}_E(\pm\sigma_\alpha)={\pi}_E( \pm i\sigma_\alpha)= \sigma_z,\quad &\text{if }\alpha=x,y.
    \end{aligned}
\end{equation}
The reader can verify that this is a well-defined representation of the Pauli group. In particular, we express this representation as 
\begin{equation}\label{eqn:decompositionPi_EPhaseDamping}
    \pi_E = \mathsf{1} \oplus \text{sgn}_{xy},
\end{equation}
where $\mathsf{1}$ is the trivial 1-dimensional representation, while $\text{sgn}_{xy}$ is a 1-dimensional representation that maps (neglecting all phases for simplicity) 
\begin{equation}\label{eqn:sgnRep}
\begin{split}
    &\text{sgn}_{xy}(\mathbb{I}_2)=\text{sgn}_{xy}(\sigma_z)=1,\\
    &\text{sgn}_{xy}(\sigma_x)=\text{sgn}_{xy}(\sigma_y)=-1.    
\end{split}
\end{equation}
Moreover, we observe that the conserved quantity $\sigma_z$ in \eqref{eqn:sigmaZconserved} implies that the corresponding $\pi_E(\sigma_z)$ in \eqref{eqn:repPhaseDamp} is trivially the identity, in accordance with our discussion in Sec.~\ref{sec:conservedQuantity}.

Finally, we remark that a Stinespring dilation of the phase damping channel can be realized also through the similar isometry 
\begin{equation}\label{eqn:VphaseDampPrime}    V'= \begin{pmatrix}
        \sqrt{1-p} & 0 \\
        -i\sqrt{p} & 0 \\
        0 & \sqrt{1-p}\\
        0 & i\sqrt{p}
    \end{pmatrix},
\end{equation}
in which we have essentially replaced $\ket{0}$ with $-i \ket{0}$ in the basis of $\mathcal{H}_E$ in \eqref{eqn:explicitConstructionDilationKraus}.
It is immediate to see that this isometry also satisfies the same covariance condition in \eqref{eqn:covariantDilationKeyl} with the environment representation in \eqref{eqn:repPhaseDamp}. We will use this isometry in the realization of a physical dilation, as the additional phases given by $\pm i$ naturally emerge in Hamiltonian evolutions.

\subsubsection{Symmetric physical dilations}
\label{sec:phaseDamSymPhysDil}
We now focus on physical dilations of time-dependent phase damping channels driven by a time-independent Hamiltonian $H_I$, as given by Eqs.~\eqref{eqn:unitaryTimeIndependentDil} and~\eqref{eqn:physDilTimeDep}. In particular, we let $p$ of the phase-damping channel depend on time, so that $\phi_{p(t)}^\text{PD}$ has the structure in \eqref{eqn:phaseDampDef} with a varying $p(t)$ for all $t\geq 0$. 
Then, the relations~\eqref{eqn:invariantState} and~\eqref{eqn:symmetryDilationParallel} for this type of dilations impose some constraints on the shape of the possible states $\ket{\psi_E}$ and Hamiltonians $H_I$ in \eqref{eqn:physDilTimeDep}. 

Our aim is to start from the environment representation $\pi_E(g)$ obtained in \eqref{eqn:repPhaseDamp} for the ``natural'' minimal dilations in \eqref{eqn:VphaseDamp} and \eqref{eqn:VphaseDampPrime}, and to find the suitable $\ket{\psi_E}$ and $H_I$ given the constraints discussed in Sec.~\ref{sec:symmetricDilTime}. The condition in \eqref{eqn:invariantState} immediately tells us what is the only possible initial state of the environment in the physical dilation, namely 
\begin{equation}\label{eqn:initialStatePhaseDamp}
    \ket{\psi_E} = \ket{1}.
\end{equation}
Indeed, $\ket{1}$ spans the subspace on which the trivial part of $\pi_E$ acts in \eqref{eqn:decompositionPi_EPhaseDamping}. 

Secondly, Eq.~\eqref{eqn:symmetryDilationParallel} states that $H_I$ must lie in the \textit{commutant} of the set (neglecting all possible trivial phases) 
\begin{equation}\label{eqn:BphaseDamp}
\begin{split}
    B &=\{\pi_S(g) \otimes \pi_E(g)\; \forall g \in G_P\}\\
    &= \{\mathbb{I}_2\otimes\mathbb{I}_2,\sigma_z\otimes\mathbb{I}_2,\sigma_x\otimes\sigma_z,\sigma_y\otimes\sigma_z\}
\end{split}
\end{equation} 
within the algebra of two-qubit observables, which is defined as 
\begin{equation}\label{eqn:commutatantDef}
    B' = \{a \in\mathcal{B}(\mathcal{H}_S\otimes\mathcal{H}_E):[a,b]=0\quad \forall b\in B\},
\end{equation}
where $\mathcal{B}(\mathcal{H}_S\otimes\mathcal{H}_E)$ is the space of bounded operators on the total Hilbert space. 
To be more precise, the commutant should be restricted to the subspace spanned by the dynamics in \eqref{eqn:sigmaParallel}, but for simplicity we do not ask for this restriction for now. 

\mar{As $B$ is a set of Pauli strings, it can be shown that its commutant is always spanned by a set of Pauli strings (see Appendix~\ref{app:pauliComm}). We then have to look simply for the Pauli strings that commute with all elements in $B$.} After some tedious algebra we find 
\begin{equation}\label{eqn:PauliCommutantDeph}
    B' = Span\{\mathbb{I}_2\otimes\mathbb{I}_2,\mathbb{I}_2\otimes\sigma_z,\sigma_z\otimes\sigma_x,\sigma_z\otimes\sigma_y\}. 
\end{equation}
According to \eqref{eqn:symmetryDilationParallel}, we have to look for a suitable $H_I$ within  $B'$\footnote{For the sake of clarity, we point out that this does not mean that any element in $B'$ provides a well-defined physical dilation of a generic $ \phi_{p(t)}^{\text{PD}}$. }.

Finally, we notice that 
\begin{equation}\label{eqn:dilationHamPhaseDamp}
    H_I=\sigma_z\otimes \sigma_x
\end{equation} 
does the job. This Hamiltonian, together with the initial state $\ket{1}_E$, generates a valid symmetric physical dilation of the phase damping channel, with isometry given by the time-dependent version of \eqref{eqn:VphaseDampPrime} and the replacement $\sqrt{1-p}\rightarrow \cos(t)$, $\sqrt{p}\rightarrow \sin(t)$. The corresponding time-dependent probability evolves as
\begin{equation}
    p(t) = \frac{1}{2}(1-\cos 2t).
\end{equation}
More generally, we can generate any time-dependent phase damping channel by employing the simple time-dependent Hamiltonian\footnote{The results presented in Sec.~\ref{sec:symmetricDilTime} can be easily extended to generic time-dependent Hamiltonians, see \cite{Cattaneo2025Dilations} for more details. In our case, $H_I(t)$ in \eqref{eqn:timeDepH_I} lies at all times in the commutant of \eqref{eqn:BphaseDamp}, so it respects the symmetry of the dilation under the action of the Pauli group.}
\begin{equation}\label{eqn:timeDepH_I}
        H_I(t)=f(t)\sigma_z\otimes \sigma_x,
\end{equation}
with $f(t)$ fixed by
\begin{equation}\label{eqn:fTvspT}
   \cos\left( 2\int_0^t ds f(s)\right) = 1-2 p(t),
\end{equation}
where $p(t)$ is a generic time-dependent probability in \eqref{eqn:phaseDampDef}.

To conclude, we notice that the dilation Hamiltonian \eqref{eqn:dilationHamPhaseDamp} is introduced, for two harmonic oscillators instead of two qubits, in the standard textbook by Nielsen and Chuang \cite{nielsenchuang} as a way to realize a phase damping channel through a Stinespring dilation. Remarkably, in this section we have derived the same Hamiltonian purely from symmetry considerations, starting from the constraints expressed in Eqs.~\eqref{eqn:invariantState} and~\eqref{eqn:symmetryDilationParallel}. We thus realize how powerful  symmetry analysis is in the construction of physical dilations of covariant channels.

\subsubsection{Freedom in the physical dilation}\label{sec:freedom}
We observe that, for the phase damping channel, we can actually realize another physical dilation with the same Hamiltonian  \eqref{eqn:dilationHamPhaseDamp} and the initial state of the environment $\ket{\psi_E}=\ket{0}$ instead of $\ket{1}$ in \eqref{eqn:initialStatePhaseDamp}. However, this is not a contradiction of the constraint in \eqref{eqn:invariantState}. In fact, changing the initial state of the environment means changing the dilation isometry, which now reads \begin{equation}
    \tilde{V}(t)=U_I(t)\ket{0}_E = \begin{pmatrix}
        -i\sin t & 0 \\
        \cos t & 0 \\
        0 & i\sin t\\
        0 & \cos t
    \end{pmatrix}.
\end{equation}
Then, the environment representation $\pi_E$ also changes, and the reader can immediately verify that, for the isometry $\tilde{V}(t)$, Eq.~\eqref{eqn:covariantDilationKeyl} implies 
\begin{equation}\label{eqn:repPhaseDampPrime}
\begin{aligned}
    &\tilde{\pi}_E(\pm\sigma_\alpha)=\tilde{\pi}_E( \pm i\sigma_\alpha)= \mathbb{I}_2,\quad &\text{if }\alpha=I,z,\\ &\tilde{\pi}_E(\pm\sigma_\alpha)=\tilde{\pi}_E( \pm i\sigma_\alpha)= -\sigma_z,\quad &\text{if }\alpha=x,y.
    \end{aligned}
\end{equation}
$\tilde{\pi}_E(g) \ket{0}_E=\ket{0}_E$, so Eq.~\eqref{eqn:invariantState} for the new dilation and $\tilde{\pi}_E$ is correctly satisfied. In general cases, however, changing the initial state of the environment does not yield a correct physical dilation, see for instance the examples in \cite{Cattaneo2025Dilations}.

Finally, following the discussion in Sec.~\ref{sec:trivialRot}, we observe that we can introduce a modified Hamiltonian 
\begin{equation}
    H_I^\text{rot} = H_I+ \mathbb{I}_S\otimes\sigma_x
\end{equation}
that produces a correct physical dilation of the phase damping channel, given that the additional environment Hamiltonian is redundant. If we write the isometry $V^\text{rot}$ for $H_I^\text{rot}$, Eq.~\eqref{eqn:covariantDilationKeyl} leads to an environment representation with a rotating phase:
\begin{equation}
    \pi^\text{rot}_E(g,t)=e^{-i\sigma_x t} \pi_E(g) e^{i\sigma_x t}\quad \forall g\in G_P.
\end{equation}
Accordingly, the symmetry constraints apply to $H_I$ and $\pi_E$, but not to $H_I^\text{rot}$ and $\pi_E^\text{rot}$.

\subsubsection{Dilations of the phase damping semigroup via collision models}\label{sec:phaseDampSemigroup}

We conclude our analysis of the phase damping channel by studying the related semigroup
\begin{equation}\label{eqn:phaseDampingSemigroup}
    \mathcal{L}^\text{PD}[\rho] = \gamma (\sigma_z \rho \sigma_z -\rho). 
\end{equation}
The semigroup generated by this Liouvillian is a phase damping channel for any $t\geq 0$ (see Appendix~\ref{appendix:propertiesPauli} for details).

Our goal is to represent this dynamical semigroup through a collision model in the limit of fast collisions. Following the general protocol outlined in \cite{Cattaneo2021,Cattaneo2022d}, we can do so by preparing unitary collisions as in \eqref{eqn:collisionUnitary} with a collision Hamiltonian that ``encodes'' all \textit{Lindblad operators} \cite{vacchini2024open} appearing in the Liouvillian. For instance, if we want to simulate a term of the Liouvillian with a Hermitian Lindblad operator $L$, we need an ancillary qubit prepared in $\ket{1}$ and a collision Hamiltonian
\begin{equation}\label{eqn:collisionHamPhaseDamp}
H_c = \nu_\text{int} L \otimes \sigma_x,  \quad \nu_\text{int}\in \mathbb{R},  
\end{equation}
 which acts on $\mathcal{H}_S\otimes\mathcal{H}_E$. $\nu_\text{int}$ is an interaction strength that scales as $\Delta t^{-\frac{1}{2}}$.
 Next, assuming for simplicity that $L$ is the only Lindblad operator of the Liouvillian, we generate a single collision through the collision unitary $U_c(\Delta t)$ and a related physical dilation of the collision map
 \begin{equation}
     \phi_{\Delta t}[\rho_S] = \Tr_E[U_c(\Delta t)\rho_S\otimes\ket{1}\!\bra{1}U_c^\dagger(\Delta t)].
 \end{equation}
 The full semigroup can then be simulated through $m$ repeated collisions with $m$ fresh ancillary qubits, 
\begin{equation}
    \exp(\mathcal{L}t)[\rho_S] = \lim_{\Delta t\rightarrow 0^+}(\phi_{\Delta t})^m[\rho_S],\quad t=m \Delta t.
\end{equation}
The finite decay rate $\gamma$ is given by
\begin{equation}
    \gamma = \lim_{\Delta t\rightarrow 0} \nu_\text{int}^2 \Delta t.
\end{equation}
We refer the readers to Appendix~\ref{app:collision} for more details.

As the only Lindblad operator of the phase damping semigroup $\mathcal{L}^\text{PD}$ is $L=\sigma_z$, we realize that the collision Hamiltonian \eqref{eqn:collisionHamPhaseDamp} we need is nothing but the dilation Hamiltonian $H_I$ in \eqref{eqn:dilationHamPhaseDamp}. Then, all the symmetry considerations for phase damping channels in Sec.~\ref{sec:phaseDamSymPhysDil} follow also for the case of phase damping semigroups.

\subsection{Depolarizing channel}
The depolarizing channel with probability $p$ is defined as
\begin{equation}\label{eqn:depChannelDef}
    \phi_p^\text{dep}[\rho_S]=(1-p)\rho_S +\frac{p}{3} \sum_{i=x,y,z}\sigma_i \rho_S \sigma_i, \quad 0\leq p\leq 1,
\end{equation}
which can be rewritten as 
\begin{equation}\label{eqn:depChannelDefBis}
       \phi_p^\text{dep}[\rho_S]=(1-\lambda)\rho_S +\frac{\lambda}{2}\mathbb{I}_2, \quad \lambda = \frac{4}{3}p,\quad 0\leq \lambda\leq \frac{4}{3}.
\end{equation}

On top of the usual covariance with respect to the Pauli group, the depolarizing channel is covariant also with respect to $SU(2)$. If we write the 2-dimensional irreducible representation (irrep) of $SU(2)$ on the space of the system qubit as 
\begin{equation}
\mu_S(\theta,\vec{n})=\exp(i\theta \vec{n}\cdot\vec{\sigma}),
\end{equation}
the channel covariance can be expressed as
\begin{equation}\label{eqn:covDepSU2}
\begin{split}
    &\mu_S(\theta,\vec{n}) \phi_p^\text{dep}[\rho_S]\mu_S^\dagger(\theta,\vec{n}) =
     \phi_p^\text{dep}[\mu_S(\theta,\vec{n})\rho_S\mu_S^\dagger(\theta,\vec{n})],    
\end{split}
\end{equation}
 for all $\theta\in\mathbb{R}$  and for all unit vectors $\vec{n}$. Eq.~\eqref{eqn:covDepSU2} follows immediately from \eqref{eqn:depChannelDefBis}.

 \subsubsection{Minimal dilation from Kraus representation}

 We follow the same lines as for the phase damping channel. We now have four Kraus operators, namely $K_0=\sqrt{1-p}\mathbb{I}_2$, $K_i = \sqrt{p/3}\sigma_i$, for $i=x,y,z$. This means that  for a minimal dilation we need the environment Hilbert space of two qubits. Using the same convention for the bases of domain and codomain as for the phase damping channel, the isometry $V$ from \eqref{eqn:explicitConstructionDilationKraus} is  
 \begin{equation}\label{eqn:isometryDep}
     V=\begin{pmatrix}
         \sqrt{1-p}& 0\\
         0& \sqrt{\frac{p}{3}}\\
         0 & -i \sqrt{\frac{p}{3}}\\
         \sqrt{\frac{p}{3}} & 0 \\
         0 & \sqrt{1-p}\\
         \sqrt{\frac{p}{3}} & 0\\
         i \sqrt{\frac{p}{3}} &0\\
         0 &-\sqrt{\frac{p}{3}}
     \end{pmatrix}.
 \end{equation}

 Next, we first focus on the covariance with respect to the Pauli group. Using Eq.~\eqref{eqn:covariantDilationKeyl}, we obtain the following representation for the space of the environment: 
 
\begin{equation}
\begin{split}\label{eqn:repDepolarizing}
    &\pi_E(\pm \mathbb{I}_2)=\pi_E(\pm i \mathbb{I}_2)=\mathbb{I}_2\otimes\mathbb{I}_2,\\
    &\pi_E(\pm \sigma_x)=\pi_E( \pm i \sigma_x)= \sigma_z\otimes\mathbb{I}_2,\\
    &\pi_E(\pm \sigma_y)=\pi_E( \pm i \sigma_y)=\mathbb{I}_2\otimes \sigma_z, \\
    &\pi_E(\pm \sigma_z)=\pi_E( \pm i \sigma_z)=\sigma_z\otimes\sigma_z.\\ 
\end{split}
\end{equation}
This representation can be written as 
\begin{equation}\label{eqn:decompPeDep}
    \pi_E = \mathsf{1}\oplus \text{sgn}_{yz} \oplus \text{sgn}_{xz} \oplus \text{sgn}_{xy},
\end{equation}
where $\text{sgn}_{ij}$ is defined similarly as in \eqref{eqn:sgnRep} (it sends the Pauli matrices $\sigma_i$ and $\sigma_j$ to $-1$, and so on).

Then, we analyze the covariance with respect to $SU(2)$. Starting from \eqref{eqn:covariantDilationKeyl}, we look for the property
\begin{equation} \label{eqn:covIsomVSU2}
    V \mu_S(\theta,\vec{n})= \mu_S(\theta,\vec{n})\otimes \mu_E(\theta,\vec{n}) V, 
\end{equation}
for some representation $\mu_E$ of $SU(2)$ on the 4-dimensional Hilbert space of the environment. Parametrizing $\mu_E$ as 
\begin{equation}\label{eqn:repSU(2)env}
    \mu_E(\theta,\vec{n})=\exp(i\theta \vec{n}\cdot\vec{J}),
\end{equation}
where $J_x$, $J_y$ and $J_z$ are 4-dimensional representations of the generators of $\mathfrak{su}(2)$, we can expand \eqref{eqn:covIsomVSU2} for each direction $x,y,z$ and for $\theta\ll 1$. Equating the terms of the same order in $\theta$, we obtain for the $x$ axis 
\begin{equation}
    \mathbb{I}_2\otimes J_x V = V \sigma_x - \sigma_x \otimes \mathbb{I}_4 V. 
\end{equation}
This equation uniquely determines $J_x$. Carrying out the same operation for all three directions, we finally get
\begin{equation}\label{eqn:generatorsu(2)}
\begin{split}
    &J_x=\begin{pmatrix}
        0 & 0 & 0 & 0 \\
        0 & 0 & 0 & 0 \\
        0 & 0 & 0 & -2i \\
        0 & 0 & 2i & 0 \\
    \end{pmatrix},\quad
    J_y=\begin{pmatrix}
        0 & 0 & 0 & 0 \\
        0 & 0 & 0 & 2i \\
        0 & 0 & 0 & 0 \\
        0 & -2i & 0 & 0 \\
    \end{pmatrix},\\
    &J_z=\begin{pmatrix}
        0 & 0 & 0 & 0 \\
        0 & 0 & -2i & 0 \\
        0 & 2i & 0 & 0 \\
        0 & 0 & 0 & 0 \\
    \end{pmatrix}.
\end{split}
\end{equation}
These operators are the generators of a 4-dimensional representation of $\mathfrak{su}(2)$ that is the direct sum of the trivial (spin-0) representation and the spin-1 representation. Therefore,
\begin{equation}\label{eqn:decompositionPiESU2}
    \mu_E = \mu^{(0)}\oplus \mu^{(1)},
\end{equation}
where $\mu^{(j)}$ denotes the spin-$j$ irrep of $SU(2)$.

\subsubsection{Symmetric physical dilations}

Next, we consider a time-dependent depolarizing channel $\phi^\text{dep}_{p(t)}$ with a probability $p(t)$, and we look for physical dilations driven by a time-independent Hamiltonian as in \eqref{eqn:unitaryTimeIndependentDil}. 

First of all, we notice that both the trivial environment representations in the decomposition of $\mu_E$ (Eq.~\eqref{eqn:decompositionPiESU2}) and $\pi_E$ (Eq.~\eqref{eqn:decompPeDep}) act on $\ket{11}$. Then, according to \eqref{eqn:invariantState} we fix the initial state of the environment
\begin{equation}\label{eqn:initialStateDep}
    \ket{\psi_E}=\ket{11}.
\end{equation}
If the trivial representations acted on different subspaces, then it would be impossible to fulfill the condition in \eqref{eqn:invariantState}, and this would be an indication that the quantum channel under consideration cannot be dilated through a time-independent Hamiltonian as in \eqref{eqn:unitaryTimeIndependentDil}.

After fixing the initial state of the environment, we now look for a suitable dilation Hamiltonian $H_I$ given the constraint in \eqref{eqn:symmetryDilationParallel}. We have now two different constraints due to the two symmetry groups and their representations in \eqref{eqn:repDepolarizing} and \eqref{eqn:repSU(2)env}. We start with the constraint associated with the Pauli group and we look for the commutant of the set
\begin{equation}\label{eqn:Bdep}
\begin{split}
    B &=\{\pi_S(g) \otimes \pi_E(g),\; \forall g \in G_P\}\\
    &=\{ \mathbb{I}_2\otimes\mathbb{I}_2\otimes\mathbb{I}_2,\sigma_z\otimes\sigma_z\otimes\sigma_z, \sigma_x\otimes\sigma_z\otimes\mathbb{I}_2,\\
    &\qquad\sigma_y\otimes\mathbb{I}_2\otimes\sigma_z\},    
\end{split}
\end{equation}
up to trivial phases. 
As shown in the \textsc{Mathematica} notebook \cite{notebook}, the commutant of $B$ contains 16 different strings of Pauli matrices.
In particular, we recognize three different strings that can independently implement the Kraus operators of the depolarizing channel, and we use them to define the dilation Hamiltonian
\begin{equation} \label{eqn:depHam_I}
    H_I = \sigma_x \otimes \mathbb{I}_2 \otimes \sigma_x + \sigma_y \otimes \sigma_x \otimes \mathbb{I}_2 + \sigma_z \otimes \sigma_x \otimes \sigma_x.
\end{equation}
The corresponding physical dilation driven by $U_I(t)=\exp(-i H_I t)$ generates a depolarizing channel for any time $t$, 
with time-dependent probability
\begin{equation}
    p(t) = \frac{1}{2}\left(1-\cos (2\sqrt{3}t)\right).
\end{equation}
Moreover, $U_I(t)$ with the initial state  \eqref{eqn:initialStateDep} generates the isometry  \eqref{eqn:isometryDep} with the replacement $\sqrt{p}\rightarrow-i\sin\sqrt{3}t$, $\sqrt{1-p}\rightarrow\cos\sqrt{3}t$.  

Once again, the symmetry constraints from the commutant of \eqref{eqn:Bdep} restricted the space of possible $H_I$, and were crucial in identifying the dilation Hamiltonian \eqref{eqn:depHam_I}. Furthermore, any depolarizing channel with generic $p(t)$ can be realized through a time-dependent version of \eqref{eqn:depHam_I}, exactly as for the phase damping channel in \eqref{eqn:timeDepH_I}. Moreover, there is a similar freedom in the definition of the dilation and related $\pi_E$ as discussed in Sec.~\ref{sec:freedom}.

Next, we analyze the symmetry properties of $H_I$ with respect to $SU(2)$. \mar{In general terms, the presence of an additional channel covariance on top of the Pauli covariance further constrains the structure of the dilation Hamiltonian in the subspace spanned by the dynamics. This is true not only for the case of $SU(2)$ in the depolarizing channel, but also for generic Pauli channels that are covariant with respect to an additional symmetry group.}

The subspace spanned by the dynamics driven by \eqref{eqn:depHam_I}, defined in \eqref{eqn:sigmaParallel}, can be constructed using Krylov subspaces, as explained in \cite{Cattaneo2025Dilations}. In particular, for the dilation considered here and initial state in \eqref{eqn:initialStateDep}, it can be shown to correspond to
\begin{equation}
    \mathcal{K}_\parallel = Span\{\ket{011},\ket{111},H_I\ket{011},H_I\ket{111}\}.
\end{equation}
In accordance with \eqref{eqn:symmetryDilationParallel}, the generators of the representation of $\mathfrak{su}(2)$ on $\mathcal{H}_S\otimes\mathcal{H}_E$, constructed through $\sigma_i$ on $\mathcal{H}_S$, and $J_i$ in \eqref{eqn:generatorsu(2)} on $\mathcal{H}_E$, commute with $H_I$ when restricted to $\mathcal{K}_\parallel$:
\begin{equation}\label{eqn:conservedSU(2)}
    \restr{[H_I,\sigma_i\otimes\mathbb{I}_4+\mathbb{I}_2\otimes J_i]}{\mathcal{K}_\parallel}=0 \text{ for all }i=x,y,z.
\end{equation}
So, these operators are conserved quantities in the subspace spanned by the dynamics. This conservation law is a constraint at the level of the dilation induced by the covariance property of the channel in \eqref{eqn:covDepSU2}.

Following \cite{Cattaneo2025Dilations}, this symmetry can be extended to the full Hilbert space $\mathcal{H}_S\otimes\mathcal{H}_E$ by defining a new interaction Hamiltonian that acts as the identity on the orthogonal complement of $\mathcal{K}_\parallel$. Indeed, $H_I$ is always decomposable into two distinct blocks, one of them acting on $\mathcal{K}_\parallel$ and the other on the orthogonal complement $\mathcal{K}_\perp$, with $\mathcal{H}_S\otimes\mathcal{H}_E=\mathcal{K}_\parallel\oplus\mathcal{K}_\perp$. The new dilation is defined by $U_I'(t)=\exp(-i H_I' t)$ and the same initial state $\ket{11}_E$, with
\begin{equation}
    H'_I = H_I\big|_{\mathcal{K}_\parallel} \oplus \mathbb{I}_{\mathcal{K}_\perp}.
\end{equation}
It can be verified that this Hamiltonian correctly generates a physical dilation of the depolarizing channel analogous to the one driven by \eqref{eqn:depHam_I} \cite{notebook}. This is consistent with the theorem in \cite{MarvianMashhad2012} that guarantees the existence of a fully symmetric physical dilation for any covariant map. However, in contrast to \cite{MarvianMashhad2012}, the symmetric dilation we have constructed is minimal and we do not need an additional degree of freedom on $\mathcal{H}_E$, thanks to the fact that the environment representations of both $SU(2)$ and $G_P$ act trivially on \eqref{eqn:initialStateDep}, as it should be for time-dependent dilations driven by a time-independent Hamiltonian.

\subsubsection{Dilations of the depolarizing semigroup via collision models}\label{sec:depSemigroup}
We conclude our discussion of the depolarizing channel by considering depolarizing semigroups driven by the Liouvillian
\begin{equation}\label{eqn:LiouvillianDep}
    \mathcal{L}^\text{dep}[\rho]=\gamma \sum_{i=x,y,z}(\sigma_i \rho \sigma_i - \rho).
\end{equation}
Our aim is to simulate this type of semigroups via fast-collision models. 

Following the discussion in Sec.~\ref{sec:phaseDampSemigroup}, we identify the three Lindblad operators $L_i=\sigma_i$ of $\mathcal{L}^\text{dep}$ and ``encode'' them in a proper collision Hamiltonian according to \eqref{eqn:collisionHamPhaseDamp}. We observe that the dilation Hamiltonian in \eqref{eqn:depHam_I} is a suitable collision Hamiltonian (see Appendix~\ref{app:collision} for more details). So, as for the phase damping channel, all the symmetry constraints discussed for dilations of time-dependent depolarizing channels are valid also for dilations of depolarizing semigroups via collision models.

\subsection{Generic \mar{single-qubit} Pauli channels}

Finally, we consider a generic single-qubit Pauli channel as defined in \eqref{eqn:PauliChannel}, with Kraus operators given by \eqref{eqn:KrausPauliChannel}. We fix $p_I$ through $p_I=1-p_x-p_y-p_z$. Then, we build a minimal Stinespring dilation following Eq.~\eqref{eqn:explicitConstructionDilationKraus}, with the environment Hilbert space of two qubits. We obtain a dilation isometry written as 
\begin{equation}
     V=\begin{pmatrix} \label{eqn:isometryGen}
         \sqrt{1-p_x-p_y-p_z}& 0\\
         0& \sqrt{p_x}\\
         0 & -i \sqrt{p_y}\\
         \sqrt{p_z} & 0 \\
         0 & \sqrt{1-p_x-p_y-p_z}\\
         \sqrt{p_x} & 0\\
         i \sqrt{p_y} &0\\
         0 &-\sqrt{p_z}
     \end{pmatrix}.
\end{equation}

The next step is to look for the environment representation $\pi_E$ of the Pauli group related to the isometry $V$ and the property in \eqref{eqn:covariantDilationKeyl}. Remarkably, for a generic $V$ in \eqref{eqn:isometryGen} we obtain the same environment representation as for the depolarizing channel, given by \eqref{eqn:repDepolarizing}: even for generic Pauli channels, $\pi_E$ is completely independent of the exact values of $p_x,p_y,p_z$. 

Inspired by this general result, we now look for a unitary dilation of generic Pauli channels that is similar to the one for the depolarizing channel. In particular, we still fix
\begin{equation}
    \ket{\psi_E}=\ket{11},
\end{equation}
which is left invariant by $\pi_E$ in \eqref{eqn:repDepolarizing}. Then, we define a dilation Hamiltonian that has the same structure as  \eqref{eqn:depHam_I} (hence, it properly commutes with $\pi_S\otimes\pi_E$), but with tunable coefficients for each Pauli string in $H_I$:
\begin{equation}
    \label{eqn:HamGen}
\begin{split}
    H_I(a_1,a_2,a_3) =& a_1  \sigma_x \otimes \mathbb{I}_2 \otimes \sigma_x + a_2\sigma_y \otimes \sigma_x \otimes \mathbb{I}_2 \\
    &+ a_3 \sigma_z \otimes \sigma_x \otimes \sigma_x,   
\end{split}
\end{equation}
where $a_1,a_2,a_3\in\mathbb{R}$ are parameters we can freely tune. For convenience, we now define
\begin{equation}\label{eqn:a1a2a3}
   \xi=  a_1^2+a_2^2+a_3^2.
\end{equation}
It can be shown that this physical dilation generates a generic Pauli channel \eqref{eqn:PauliChannel} with time-dependent probabilities 
\begin{equation}
\begin{split}
    &p_x(t) = a_1^2 \frac{\sin^2(\sqrt{\xi}t)}{\xi},\\
    &p_y(t) = a_2^2 \frac{\sin^2(\sqrt{\xi}t)}{\xi},\\
    &p_z(t) = a_3^2 \frac{\sin^2(\sqrt{\xi}t)}{\xi}, \\
    &p_I(t) = 1-\sin^2(\sqrt{\xi}t).    
\end{split}
\end{equation}
\mar{Exploring more general symmetric structures of the Hamiltonian and/or allowing the coefficients $a_i$ to depend on time, we can generate any possible time-dependent single-qubit Pauli channel, as we will show for the more general $n$-qubit case in the next section.}

Finally, we address the simulation of a generic Pauli semigroup in \eqref{eqn:PauliSemigroup} through fast-collision models, following the lines of the discussion in Secs.~\ref{sec:phaseDampSemigroup} and~\ref{sec:depSemigroup}. As for the phase damping and depolarizing channels, we quickly realize that the Hamiltonian in \eqref{eqn:HamGen}, without any constraints on the coefficients $a_i\in\mathbb{R}$, is a suitable collision Hamiltonian, as discussed in Appendix~\ref{app:collision}. Then, once again the same symmetry properties as for generic Pauli channels apply to collision Hamiltonians for generic Pauli semigroups. 

\mar{
\section{Symmetric dilations of multi-qubit Pauli channels}\label{sec:resultsMulti}
\subsection{Minimal dilation from Kraus representation}\label{sec:minKrauMulti}
Once again, we first construct the general isometry $V$ of a generic $n$-qubit Pauli channel in \eqref{eqn:nQubitPauliChannel} by using \eqref{eqn:explicitConstructionDilationKraus}:
\begin{equation}\label{eqn:isometryMultiQubit}
    V \ket{\varphi_S} = \sum_{a\in A} \sqrt{p_a} P_a \ket{\varphi_S}\otimes\ket{e_a},
\end{equation}
where $\{\ket{e_a}\}_{a\in A}$ is a basis of the environment Hilbert space, whose dimension is determined by the support of the channel $A$, i.e., by its Kraus rank. Next, we look for a representation of $G_P^{(n)}$ on the space of the environment satisfying \eqref{eqn:covariantDilationKeyl} with the above $V$ and $\pi_S$ being the defining representation of $G_P^{(n)}$. 
}

\mar{
Given a generic $P_b\in G_P^{(n)}$, we introduce the representation
\begin{equation}
    \label{eqn:repEnvNqubit}
    \begin{split}
    \pi_E(\pm P_b)=\pi_E(\pm i P_b) &=\sum_{a\in A}(-1)^{\omega(a,b)}\ket{e_a}\!\bra{e_a}.
    \end{split}
\end{equation}
We can decompose this representation as 
\begin{equation}
    \pi_E= \bigoplus_{a\in A}\chi_a,
\end{equation}
where each
\begin{equation}
    \chi_a(\pm P_b)=\chi_a(\pm i P_b) = (-1)^{\omega(a,b)}
\end{equation}
is a one-dimensional representation (character) of the Pauli group.  It is easy to check that this is a valid environment representation for $V$ in \eqref{eqn:isometryMultiQubit}:
\begin{equation}
\begin{split}
P_b\otimes \pi_E(P_b) V &= \sum_{a,a'\in A}\sqrt{p_a}(-1)^{\omega(a,b)}\delta_{aa'} P_b P_a\otimes\ket{e_a}\\
    &= \sum_{a\in A}\sqrt{p_a} P_a P_b \otimes \ket{e_a}=V P_b,\\    
\end{split}
\end{equation}
recovering \eqref{eqn:covariantDilationKeyl}.
}

\mar{
Remarkably, this environment representation does not depend on the coefficients $p_a$, but only on the support $A$ of the channel in \eqref{eqn:nQubitPauliChannel}. This explains why the environment representation for the single-qubit depolarizing channel is the same as for
 generic single-qubit Pauli channels (Eq.~\eqref{eqn:repDepolarizing}). In particular, we observe that restricting \eqref{eqn:repEnvNqubit} to a single qubit with $A=\{(0|0),(0|1)\}$ we obtain the representation for the phase damping channel in \eqref{eqn:repPhaseDamp}. Similarly, if the single-qubit support contains all the four possible binary strings, then we recover the representation in \eqref{eqn:repDepolarizing}. Moreover, the only state of the environment that satisfies the constraint in \eqref{eqn:invariantState} is $\ket{e_0}$,
 i.e., the state associated with the trivial representation $\chi_0$. If the environment is composed of qubits, using the same convention as for single-qubit channels, $\ket{e_0}$ is given by 
 \begin{equation}\label{eqn:invariantStateMultiQubit}
     \ket{\psi_E} = \ket{e_0} =\ket{11\ldots 1}.
 \end{equation}
}
\mar{
\subsection{Symmetric physical dilations}
Next, starting from the symmetry rules imposed by \eqref{eqn:symmetryDilationParallel} with the representation in \eqref{eqn:repEnvNqubit}, we can construct the most generic symmetric dilation Hamiltonian driving time-dependent dilations as in \eqref{eqn:unitaryTimeIndependentDil}. Indeed, let us introduce the unitary operator 
\begin{equation}
    \label{eqn:opC}
    C = \sum_{a\in A} P_a\otimes \ket{e_a}\!\bra{e_a},
\end{equation}
which conjugates the representation of $G_P^{(n)}$ on the total Hilbert space as follows:
\begin{equation}
\begin{split}
    C^\dagger P_b\otimes \pi_E(P_b) C &=\sum_{a\in A} (-1)^{\omega(a,b)} P_a P_b P_a \otimes \ket{e_a}\!\bra{e_a} \\
    &= P_b\otimes\mathbb{I}_E. 
\end{split}
\end{equation}
Then, looking for a generic $H_I$ such that 
\begin{equation}\label{eqn:commutHInqubit}
    [H_I, P_b\otimes \pi_E(P_b)]=0\quad \forall\, P_b\in G_P^{(n)}
\end{equation}
is equivalent to the constraint
\begin{equation}
    [C^\dagger H_I C, P_b\otimes\mathbb{I}_E]=0 \quad \forall \,P_b\in G_P^{(n)},
\end{equation}
which in turn implies 
\begin{equation}\label{eqn:constraintMultiQubit}
C^\dagger H_I C = \mathbb{I}_S \otimes O,    
\end{equation}
for a generic Hermitian $O\in\mathcal{B}(\mathcal{H}_E)$, due to Schur's lemma and the fact that the defining representation of $G_P^{(n)}$ is irreducible.}

\mar{To conclude, a dilation Hamiltonian $H_I$ commutes with all elements in the tensor product representation of $G_P^{(n)}$ in \eqref{eqn:commutHInqubit} if and only if it is of the form 
\begin{equation}
    \label{eqn:genericHInqubit}
    H_I = C\, \mathbb{I}_S \otimes O \, C^\dagger,\text{ for some } \,O\in\mathcal{B}(\mathcal{H}_E)\text{ with }O^\dagger = O.
\end{equation}
Then, if $r_K$ is the dimension of the Hilbert space of the environment dilation (or equivalently the size of the support $A$),  the commutant of $\{P_b\otimes \pi_E(P_b)\;\forall \, P_b\in G_P^{(n)}\}$ is spanned by $r_K^2$ linearly independent Hermitian matrices. This result explains why we found that exactly 4 matrices span the commutant for the phase damping channel in \eqref{eqn:PauliCommutantDeph}, while 16 matrices span the commutant of the set of matrices \eqref{eqn:Bdep} for depolarizing and generic single-qubit Pauli channels. Moreover, in the \textsc{Mathematica} notebook we show that the dilation Hamiltonian \eqref{eqn:HamGen} for generic single-qubit Pauli channels can be recovered through a specific operator $O$ and the structure in \eqref{eqn:genericHInqubit} \cite{notebook}.
}

\mar{
Interestingly, the evolution of the dilation with initial state of the environment $\ket{e_0}$ can be expressed in a particularly simple form: 
\begin{equation}\label{eqn:quantumWalk}
    e^{-i H_I t} \ket{\varphi_S}\otimes\ket{e_0} =\sum_{a\in A} \bra{e_a}e^{-i O t}\ket{e_0} P_a \ket{\varphi_S} \otimes\ket{e_a}.
\end{equation}
We can interpret this evolution as a continuous-time quantum walk on the graph of environment states $\{\ket{e_a}\}_{a\in A}$ driven by the complex Hamiltonian $O$, with the conditional application of the Pauli string $P_a$ on the initial state of the system $\ket{\varphi_S}$ for each environment state $\ket{e_a}$ reached by the quantum walk. 
}

\mar{
\subsubsection{Time-dependent Hamiltonians and generic time-dependent Pauli channels}
Finally, we observe that the constraint in \eqref{eqn:constraintMultiQubit} holds at any time $t$. So, the structure in \eqref{eqn:genericHInqubit} with a time-dependent $O(t)$ generates any symmetric time-dependent dilation Hamiltonian $H_I(t)$. Then, suppose we aim to dilate a generic multi-qubit Pauli channel with smooth time-dependent coefficients $p_a(t)$ and, without losing generality, we need to fix $p_a(0)=\delta_{a 0}$:
\begin{equation}\label{eqn:timeDepMultiQPaul}
    \phi(t)[\rho_S] = \sum_{a\in A}p_a(t) P_a \rho_S P_a.
\end{equation}
We introduce the time-dependent environment state
\begin{equation}
    \ket{\psi_E(t)}= \sum_{a\in A} \sqrt{p_a(t)}\ket{e_a},
\end{equation}
and the time-dependent operator 
\begin{equation}
    O(t) = i\left(\ket{\frac{d}{dt}\psi_E(t)}\bra{\psi_E(t)}-\ket{\psi_E(t)}\bra{\frac{d}{dt}\psi_E(t)}\right).
\end{equation}
Using $\braket{\frac{d}{dt}\psi_E(t)|\psi_E(t)}=0$, we observe $O(t)\ket{\psi_E(t)}=i\frac{d}{dt}\ket{\psi_E(t)}$, therefore
\begin{equation}
    U_E(t)\ket{e_0} = \mathcal{T}e^{-i\int_0^t ds \,O(s)}\ket{e_0}= \ket{\psi_E(t)}.
\end{equation}
Then, 
\begin{equation}
    U_I(t) = \mathcal{T}e^{-i \int_0^t ds\, C \,\mathbb{I}_S\otimes O(s) \,C^\dagger} = C\, \mathbb{I}_S\otimes U_E(t)\, C^\dagger,
\end{equation}
and 
\begin{equation}
    \phi(t)[\rho_S] = \Tr_E[U_I(t)\rho_S\otimes\ket{e_0}\!\bra{e_0}U_I^\dagger(t)].
\end{equation}
To conclude, the time-dependent dilation Hamiltonian $H_I(t)=C\, \mathbb{I}_S\otimes O(t) \, C^\dagger $ is symmetric with respect to the tensor product representation of the Pauli group $P_b\otimes \pi_E(P_b)$, and it generates any possible time-dependent multi-qubit Pauli channel in \eqref{eqn:timeDepMultiQPaul} when the environment starts from the invariant initial state $\ket{e_0}$ in \eqref{eqn:invariantStateMultiQubit}.
}

\mar{
\subsection{Dilations of generic multi-qubit Pauli semigroups via collision models}
A rapid calculation shows that the Hamiltonian \eqref{eqn:genericHInqubit} can also generate any possible multi-qubit Pauli semigroup in \eqref{eqn:nQubitPauliSemigroup}, following the same steps as for single-qubit semigroups discussed in Appendix~\ref{app:collision}. More specifically, up to the rescaling through a global constant, a collision model with Hamiltonian \eqref{eqn:genericHInqubit} in the fast-collision limit generates the quantum dynamical semigroup driven by the Liouvillian
\begin{equation}
    \mathcal{L}[\rho_S] = \sum_{a\in A} \abs{\braket{e_a|O|e_0}}^2 (P_a\rho_S P_a-\rho_S).
\end{equation}
Given the freedom we have in tuning the off-diagonal elements of $O$, we can easily reproduce any multi-qubit Pauli semigroup.
}

\section{Conclusions}\label{sec:conclusions}

In this paper, we have addressed the problem of identifying the symmetry properties of physical dilations of \mar{multi-qubit} Pauli channels and semigroups, starting from their fundamental covariance property with respect to the Pauli group\mar{, and showed how these can be exploited to construct suitable dilations of generic time-dependent Pauli channels}. 

For \mar{single-qubit phase damping and depolarizing channels, and generic single- and multi-qubit Pauli channels,} we have constructed the minimal dilation isometry $V$ starting from the Kraus operators, using the paradigmatic connection between Kraus and Stinespring representations of quantum channels expressed in \eqref{eqn:explicitConstructionDilationKraus}. Then, we have derived the corresponding representation $\pi_E$ of the Pauli group on the Hilbert space of the environment whose existence is guaranteed by a theorem by Scutaru, Keyl, and Werner \cite{Scutaru1979,Keyl1999}, which is summarized in Eq.~\eqref{eqn:covariantDilationKeyl}. We have noticed that $\pi_E$ has a very \mar{elegant and general form for multi-qubit channels, which depends only on the support of the channel but not on the channel coefficients (Eq.~\eqref{eqn:repEnvNqubit}). This form can be expressed in a very compact way for single-qubit depolarizing and generic channels (Eq.~\eqref{eqn:repDepolarizing}), and is even simpler for phase damping channels due to the fact that they have two Kraus operators only (Eq.~\eqref{eqn:repPhaseDamp}),} as in this case the environment is just a single qubit. For the depolarizing channel, we have also obtained the representation $\mu_E$ of $SU(2)$ on the space of the environment, given by \eqref{eqn:repSU(2)env}. These results, which can be found in Secs.~\ref{sec:results} and~\ref{sec:resultsMulti}, are of  general interest for the formal characterization of Stinespring dilations of \textit{any} \mar{multi-qubit} Pauli channel, as they shed light on a general covariance property of the dilation inherited from the covariance property of the channel. 

Our analysis was not limited to a formal characterization of the covariance property at the level of the dilation. Indeed, we have also explored how this covariance shapes the structure of physical dilations of Pauli channels driven by a Hamiltonian, or of Pauli semigroups simulated by collision models in the limit of fast collisions. First, we complemented some recent results on this type of dilations \cite{Cattaneo2025Dilations} in the discussion in Sec.~\ref{sec:symmetricDilTime}. More specifically, we have explored the role of covariant channels with associated strongly conserved quantities (Sec.~\ref{sec:conservedQuantity}), and we have shown that the group representation on the environment may have a trivial rotating phase (Sec.~\ref{sec:trivialRot}). Then, we have applied these results to Pauli channels and semigroups. 

By exploiting the constraints on this type of physical dilations for covariant channels expressed by Eqs.~\eqref{eqn:invariantState} and~\eqref{eqn:symmetryDilationParallel}, we have \mar{obtained the most general form that a symmetric Hamiltonian for physical dilations of generic multi-qubit Pauli channels must respect (Eq.~\eqref{eqn:genericHInqubit}). By allowing this Hamiltonian to depend on time, this structure and an initial invariant state of the environment can generate any possible time-dependent Pauli channel. This result captures the more specific case of single-qubit Pauli channels, and we} explicitly constructed dilation Hamiltonians for the phase damping (Eq.~\eqref{eqn:dilationHamPhaseDamp}), depolarizing (Eq.~\eqref{eqn:depHam_I}), and generic \mar{single-qubit} Pauli channels (Eq.~\eqref{eqn:HamGen}).  Moreover, we have noticed that the same Hamiltonians can be used as collision Hamiltonians of fast-collision models for the simulation of the corresponding \mar{single- or multi-qubit} Pauli semigroups.  

In all cases, the symmetry constraints helped us find suitable dilations and identify the initial states of the environment. Remarkably, the unitary dilations we have derived \textit{must} have an associated set of conserved quantities that are representations of the Pauli group, namely Eq.~\eqref{eqn:BphaseDamp} for phase damping channels and Eq.~\eqref{eqn:Bdep} for depolarizing and generic \mar{single-qubit} Pauli channels. For the depolarizing channel, we have also found the associated set of conserved quantities on the subspace spanned by the dynamics due to the covariance with respect to $SU(2)$, given by Eq.~\eqref{eqn:conservedSU(2)}. \mar{This additional covariance further constrains the structure of the corresponding dilation Hamiltonian.}

Future work may include the case of \textit{Pauli component erasing} channels \cite{DeLeon2022}. Moreover, a similar analysis may be performed for the more general case of \textit{Weyl channels} that act on qudits and are covariant with respect to the finite group generated by Weyl operators \cite{Datta2006,Amosov2007,Nathanson2007,Siudzinska2018,Siudzinska2020}, and for their multipartite extensions \cite{Basile2024Weyl,Chruscinski2025}. \mar{In this case, we expect a construction similar to the one presented in Sec.~\ref{sec:resultsMulti}, with the key difference that Weyl operators are unitary but non-Hermitian.} Finally, we would like to point out that our study of collision Hamiltonians for Pauli semigroups, including the discussion in Appendix~\ref{app:collision}, may pave the way for a general theory of \textit{minimal} dilations of quantum dynamical semigroups through collision models in the limit of infinitesimal timestep, which, to the best of our knowledge and understanding, has not been formalized yet.

\mar{
\section*{Disclaimer on the use of AI tools}
AI tools, and specifically Claude Opus 5.5, were employed for some analytical calculations that we have presented in the revised version of this manuscript, while no AI tool was used to produce the first version of our work. In particular, we obtained the generalization of the single-qubit results for generic Pauli channels to $n$ qubits (mostly Section~\ref{sec:resultsMulti}) with the assistance of Opus 5.5. All results were checked and validated by the author, who takes full responsibility for the content.  
}

\section*{Acknowledgments}
We would like to thank Maciej Malinowski for sending us a report generated through a large-language-model (LLM) pointing out the possibility of extending some of our single-qubit results to $n$ qubits. In particular, this report got the same environment representation for $n$ qubits that we have presented in Section~\ref{sec:minKrauMulti}, which we then independently derived with the assistance of Claude Opus 5.5. The suggestions in the rest of the report instead differ from the results presented here.

We would also like to thank José Alfredo de León and Alejandro Fonseca for inspiring this work and for interesting discussions, as well as Nicola Pranzini for a critical reading of the original draft. We acknowledge funding from the Research Council of Finland
through the Finnish Quantum Flagship Project No. 358878, from COQUSY Project No. PID2022-140506NB-C21 funded by MCIN/AEI/10.13039/501100011033, and from the Juan de la Cierva fellowship JDC2024-054027-I funded by MI-CIU/AEI/10.13039/501100011033.

\appendix

\section{Properties of Pauli channels}\label{appendix:propertiesPauli}

\subsection{Representation on the Bloch sphere}
We first prove Eq.~\eqref{eqn:PaulionBloch} using the properties of Pauli matrices, and in particular 
\begin{equation}
    \sigma_i \sigma_j \sigma_i = 2\sigma_i \delta_{ij} -\sigma_j.
\end{equation} Given a Pauli channel $\phi$ and $\rho_S$ from \eqref{eqn:blochSphere},
\begin{equation}
    \begin{split}
        \phi[\rho_S]=&p_I \frac{\mathbb{I}_2}{2}+ p_I \sum_i \frac{r_i}{2}\sigma_i +\sum_i p_i \frac{\mathbb{I}_2}{2} \\
        &+ \sum_{i,j}  \frac{p_i r_j}{2} \sigma_i \sigma_j \sigma_i\\
        =&\frac{\mathbb{I}_2}{2}+ p_I \sum_i \frac{r_i}{2}\sigma_i + \sum_{j}  \frac{r_j}{2} \left(2p_j-\sum_i p_i\right)\sigma_j 
        \\
        =&\frac{\mathbb{I}_2}{2}+ \sum_i \frac{\lambda_i r_i}{2} \sigma_i,
    \end{split}
\end{equation}
where $\lambda_i$ are given by \eqref{eqn:PauliScalingComponents}.

\subsection{Covariance property}
Next, the covariance property of Pauli channels is simply proven by observing that
\begin{equation}
    \begin{split}
        \phi[\sigma_i \rho_S \sigma_i ]&=p_I \sigma_i \rho_S \sigma_i + \sum_j p_j \sigma_j\sigma_i \rho_S \sigma_i \sigma_j\\
        & = p_I \sigma_i \rho_S \sigma_i + \sum_j p_j \sigma_i \sigma_j \rho_S  \sigma_j \sigma_i =\sigma_i\phi[\rho_S]\sigma_i.
    \end{split}
\end{equation}
In the above equation, we have used the anticommutative properties of the Pauli matrices:
\begin{equation}
    \sigma_i \sigma_j = \eta_{ij} \sigma_j  \sigma_i, \qquad \eta_{ij} = \begin{cases}
        1 & \text{if }i=j,\\
        -1 & \text{if }i\neq j.\\
    \end{cases}
\end{equation}
\mar{For multi-qubit Pauli channels the proof is analogous:
\begin{equation}
\begin{split}
    &\sum_{a} p_a P_b P_a \rho_S P_a P_b =   \sum_{a} p_a (-1)^{2\omega(a,b)}  P_a P_b \rho_S P_b P_a \\ &=   \sum_{a} p_a  P_a P_b \rho_S P_b P_a.
\end{split}
\end{equation}
}

\subsection{Pauli semigroups}
Finally, we prove that a Pauli semigroup as given by \eqref{eqn:PauliSemigroup} is a Pauli channel for every time $t$. We notice  that
\begin{equation}
    \mathcal{L}[\mathbb{I}_2] = 0,\quad \mathcal{L}[\sigma_i] = -2 \sum_{j\neq i} \gamma_j \sigma_i. 
\end{equation}
Therefore, using Eq.~\eqref{eqn:blochSphere} we obtain 
\begin{equation}
    \begin{split}
        e^{\mathcal{L}t}[\rho_S] = \frac{1}{2}\left(\mathbb{I}_2 + \sum_i r_i e^{-2\sum_{j\neq i}\gamma_j t}\sigma_i \right),
    \end{split}
\end{equation}
and we immediately realize that a Pauli semigroup at time $t$ is a Pauli channel that scales vectors on the Bloch sphere as in \eqref{eqn:PauliVectorRescaling}, with coefficients 
\begin{equation}
    \lambda_i(t) = e^{-2\sum_{j\neq i}\gamma_j t}.
\end{equation}
Analogously, we can express it as a Pauli channel through the Kraus decomposition in \eqref{eqn:PauliChannel} and
\begin{equation}
\begin{split}
    p_I(t)&=\frac{1}{4}\left(1+\sum_i e^{-2\sum_{j\neq i}\gamma_jt}\right),\\
    p_i(t)&=\frac{1}{4}\left(1+e^{-2\sum_{j\neq i}\gamma_jt} -\sum_{k\neq i} e^{-2\sum_{j\neq k}\gamma_j t}\right).
    \end{split}
\end{equation}

More specifically, for the phase damping semigroup \eqref{eqn:phaseDampingSemigroup} we have $\gamma_z=\gamma$, $\gamma_x=\gamma_y=0$. Hence,
\begin{equation}
    \lambda_z(t)=1,\quad \lambda_x(t)=\lambda_y(t)=e^{-2\gamma t},
\end{equation}
and
\begin{equation}
\begin{split}
    &p_I(t)=\frac{1}{2}\left(1+e^{-2\gamma t}\right),\\
    &p_z(t)=\frac{1}{2}\left(1-e^{-2\gamma t}\right),\\
    &p_x(t)=p_y(t)=0. 
    \end{split}
\end{equation}

Finally, the depolarizing semigroup \eqref{eqn:LiouvillianDep} is defined by $\gamma_x=\gamma_y=\gamma_z=\gamma$. Therefore
\begin{equation}
    \lambda_x(t)=\lambda_y(t)=\lambda_z(t)=e^{-4\gamma t},
\end{equation}
and
\begin{equation}
\begin{split}
    &p_I(t)=\frac{1}{4}\left(1+3e^{-4\gamma t}\right),\\
    &p_i(t)=\frac{1}{4}\left(1-e^{-4\gamma t}\right)\text{ for }i=x,y,z.
\end{split}
\end{equation}

\mar{To conclude, we prove that an $n$-qubit Pauli dynamical semigroup is also an $n$-qubit Pauli channel: let us first introduce the superoperator 
\begin{equation}
    \mathcal{P}_a[\rho_S]= P_a \rho_S P_a.
\end{equation}
It is easy to verify that $\mathcal{P}_a^2=\text{id}$ and $\mathcal{P}_a\mathcal{P}_b=\mathcal{P}_b\mathcal{P}_a$. Then, for an $n$-qubit Pauli dynamical semigroup driven by the Liouvillian $\mathcal{L}$,
\begin{equation}
    e^{\mathcal{L}t} = \prod_{a\neq 0} e^{\gamma_a t (\mathcal{P}_a-\text{id})} = \prod_{a\neq 0} ((1-q_a)\text{id}+ q_a \mathcal{P}_a),
\end{equation}
with $q_a = (1-e^{-2\gamma_a t})/2$. $\exp(\mathcal{L}t)$ is therefore a composition of several $n$-qubit Pauli channels, which is in turn a Pauli channel because $\mathcal{P}_a\mathcal{P}_b=\mathcal{P}_{a\oplus b}$,
where $a\oplus b$ is the bitwise sum mod 2:
\begin{equation}
    a\oplus b = (a^x \oplus b^x| a^z \oplus b^z),
\end{equation}
and $(a^x\oplus b^x)_j = a_j^x+b_j^x\, \text{ mod }2$.
}

\mar{
\section{Commutant of a set of Pauli strings}\label{app:pauliComm}
For the sake of generality, we focus on generic $n$-qubit Pauli strings. Given a set of Pauli strings $B=\{P_a,\, a\in A\}$ with some support $A$, our aim is to identify the commutant $B'$ of operators that commute with all elements in $B$. Let us write such an operator $X$ in the basis of Pauli strings: 
\begin{equation}
    X = \sum_s c_s P_s,
\end{equation}
with some coefficients $c_s$. Then, the property $[X,P_a]=0$ for a generic $P_a$ in $B$ leads us to 
\begin{equation}
    \sum_s c_s P_a P_s P_a = \sum_s c_s P_s (-1)^{\omega(a,s)} = \sum_s c_s P_s. 
\end{equation}
As the $P_s$ form a basis of the space of $n$-qubit operators, the above expression implies that, for each $s$ in the decomposition of $X$, either $\omega(a,s)$ is even or $c_s=0$. So, all Pauli strings in the decomposition of $X$ must commute with every $P_a$ in $B$. In other words, the commutant $B'$ is spanned by the Pauli strings that commute with the whole $B$.
}

\section{Collision Hamiltonian for Pauli semigroups}\label{app:collision}

We follow the standard derivation of the Lindblad equation generated by a fast-collision model \cite{Cattaneo2022d}. We define the collision Hamiltonian as 
\begin{equation}\label{eqn:collisionHamGen}
    H_c= \nu_\text{int}\sum_i L_i \otimes B_i, \quad \nu_\text{int}\in\mathbb{R},
\end{equation}
where $L_i$ (respectively $B_i$) are operators acting on $\mathcal{H}_S$ ($\mathcal{H}_E$), and $\nu_\text{int}$ is an interaction strength that goes like $\Delta t ^{-\frac{1}{2}}$. Then, the quantum map associated with a single collision driven by \eqref{eqn:collisionHamGen} and initial state $\ket{\psi_E}$ is given by \cite{Cattaneo2022d}
\begin{equation}
\begin{split}
    \phi_{\Delta t}[\rho_S] =& \rho_S + \\
    &\Delta t^2 \nu_\text{int}^2 \sum_{i,j} c_{ij}\left(L_j \rho_S L_i^\dagger-\frac{1}{2}\{L_i^\dagger L_j,\rho_S\}\right), 
\end{split}
   \end{equation}
   with 
   \begin{equation}\label{eqn:coefficientsCollision}
        c_{ij}= \Tr_E[B_{i}^\dagger B_j \ket{\psi_E}\!\bra{\psi_E}],
   \end{equation}
   and under the assumption 
   \begin{equation}\label{eqn:condColl}
       \Tr_E[B_i \ket{\psi_E}\!\bra{\psi_E}]=0. 
   \end{equation}
Interpreting $\phi_{\Delta t}[\rho_S(t)]$ as $\rho_S(t+\Delta t)$, the Liouvillian associated with this collision model in the limit of fast collisions is then defined by
\begin{equation}
    \mathcal{L}_c[\rho_S] = \lim_{\Delta t\rightarrow 0}\frac{\phi_{\Delta t}[\rho_S]-\rho_S}{\Delta t},
\end{equation}
which is proportional to the finite time-independent constant 
\begin{equation}
    \zeta = \lim_{\Delta t\rightarrow 0} \nu_\text{int}^2 \Delta t.
\end{equation}

To generate a \mar{single-qubit} Pauli semigroup \eqref{eqn:PauliSemigroup}, we need a collision Hamiltonian \eqref{eqn:collisionHamGen} with
\begin{equation}
    L_i = a_i \sigma_i, \quad i=x,y,z, \quad a_i\in\mathbb{R}.
\end{equation}
Inspired by the dilation Hamiltonian of a depolarizing or generic Pauli channel in Eqs.~\eqref{eqn:depHam_I} and~\eqref{eqn:HamGen}, we then choose 
\begin{equation} \label{eqn:Bop}
    B_x = \mathbb{I}_2\otimes\sigma_x, \quad B_y = \sigma_x\otimes \mathbb{I}_2, \quad B_z=\sigma_x\otimes\sigma_x. 
\end{equation}
This collision Hamiltonian is essentially Eq.~\eqref{eqn:HamGen} up to the rescaling through the interaction strength $\nu_\text{int}$.

We observe that the operators  \eqref{eqn:Bop} together with $\ket{\psi_E}=\ket{11}$ satisfy the condition  \eqref{eqn:condColl}, and the associated coefficients \eqref{eqn:coefficientsCollision} are  $c_{ij}= \delta_{ij}$. Then, this collision model generates a generic Pauli semigroup \eqref{eqn:PauliSemigroup} with decay rates
\begin{equation}
    \gamma_i = \zeta a_i^2.
\end{equation}
The depolarizing semigroup is recovered in the simple case $a_x=a_y=a_z$. 

We observe that in the general simulation protocol in \cite{Cattaneo2021} a different ancillary qubit for each Lindblad operator is required. In the present discussion, we instead just required a different orthonormal basis state of the environment for each Lindblad operator. Then, this procedure may pave the way for a rigorous theory of minimal dilations of quantum dynamical semigroups via fast-collision models.

\bibliography{biblio}

@misc{notebook,
  author       = {Marco Cattaneo},
  title        = {{\fontfamily{qcr}\selectfont Pauli\_dilations}: A notebook for the characterization of symmetries in physical dilations of {P}auli channels},
  year         = {2026},
publisher = {GitHub},
journal = {GitHub repository},
  howpublished = {\url{https://github.com/MarcoCattaneo/Pauli_dilations}}
}

@article{Farhi1998,
author = {Farhi, Edward and Gutmann, Sam},
journal = {Phys. Rev. A},
number = {58},
pages = {915},
title = {{Quantum computation and decision trees}},
url = {https://doi.org/10.1103/PhysRevA.58.915},
volume = {58},
year = {1998}
}

@misc{dion2024efficiently,
  title={Efficiently manipulating Pauli strings with PauliArray},
  author={Dion, Maxime and Belabbas, Tania and Bastien, Nolan},
eprint = {2405.19287},
url = {https://arxiv.org/abs/2405.19287},
primaryClass = {quant-ph},
  year={2024}
}

@article{Basile2024Pauli,
author = {Basile, Tom{\'{a}}s and Pineda, Carlos},
doi = {10.1371/journal.pone.0297210},
editor = {Barbieri, Marco},
isbn = {1111111111},
issn = {1932-6203},
journal = {PLOS ONE},
month = {apr},
number = {4},
pages = {e0297210},
pmid = {38598439},
title = {{Quantum simulation of Pauli channels and dynamical maps: Algorithm and implementation}},
url = {https://dx.plos.org/10.1371/journal.pone.0297210},
volume = {19},
year = {2024}
}

@article{Chruscinski2024,
author = {Chru{\'{s}}ci{\'{n}}ski, Dariusz and Kimura, Gen and Mukhamedov, Farrukh},
doi = {10.1088/1751-8121/ad3c82},
issn = {1751-8113},
journal = {J. Phys. A: Math. Theor.},
month = {may},
number = {18},
pages = {185302},
title = {{Universal constraint for relaxation rates of semigroups of qubit Schwarz maps}},
url = {https://iopscience.iop.org/article/10.1088/1751-8121/ad3c82},
volume = {57},
year = {2024}
}

@article{Basile2024Weyl,
author = {Basile, Tom{\'{a}}s and de Leon, Jose Alfredo and Fonseca, Alejandro and Leyvraz, Fran{\c{c}}ois and Pineda, Carlos},
doi = {10.1103/PhysRevA.109.032607},
issn = {2469-9926},
journal = {Phys. Rev. A},
month = {mar},
number = {3},
pages = {032607},
title = {{Weyl channels for multipartite systems}},
url = {https://link.aps.org/doi/10.1103/PhysRevA.109.032607},
volume = {109},
year = {2024}
}

@article{Acevedo2026,
author = {Acevedo, Alberto and Falcó, Antonio},
journal = {J. Phys. A: Math. Theor.},
title = {{Geometric Complexity of Quantum Channels via Unitary Dilations}},
volume = {59},
pages={205303},
url = {https://iopscience.iop.org/article/10.1088/1751-8121/ae6add},
year = {2026}
}

@article{Jagadish2020,
author = {Jagadish, Vinayak and Srikanth, R. and Petruccione, Francesco},
doi = {10.1103/PhysRevA.101.062304},
issn = {24699934},
journal = {Phys. Rev. A},
number = {6},
pages = {062304},
publisher = {American Physical Society},
title = {{Convex combinations of pauli semigroups: Geometry, measure, and an application}},
volume = {101},
year = {2020}
}

@article{Lancien2024,
author = {Lancien, C{\'{e}}cilia and Winter, Andreas},
doi = {10.22331/q-2024-04-30-1320},
issn = {2521-327X},
journal = {Quantum},
month = {apr},
number = {1},
pages = {1320},
title = {{Approximating quantum channels by completely positive maps with small Kraus rank}},
url = {https://quantum-journal.org/papers/q-2024-04-30-1320/},
volume = {8},
year = {2024}
}

@article{Memarzadeh2022,
author = {Memarzadeh, Laleh and Sanders, Barry C.},
doi = {10.1103/PhysRevResearch.4.033206},
issn = {2643-1564},
journal = {Phys. Rev. Research},
month = {sep},
number = {3},
pages = {033206},
publisher = {American Physical Society},
title = {{Group-covariant extreme and quasiextreme channels}},
url = {https://link.aps.org/doi/10.1103/PhysRevResearch.4.033206},
volume = {4},
year = {2022}
}

@article{Chruscinski2025,
author = {Chru{\'{s}}ci{\'{n}}ski, Dariusz and Bhattacharya, Bihalan and Patra, Saikat},
doi = {10.3390/sym17060943},
issn = {2073-8994},
journal = {Symmetry},
month = {jun},
number = {6},
pages = {943},
title = {{Symmetries of Multipartite Weyl Quantum Channels}},
url = {https://www.mdpi.com/2073-8994/17/6/943},
volume = {17},
year = {2025}
}

@article{Nathanson2007,
author = {Nathanson, Michael and Ruskai, Mary Beth},
doi = {10.1088/1751-8113/40/28/S22},
issn = {1751-8113},
journal = {J. Phys. A: Math. Theor.},
month = {jul},
number = {28},
pages = {8171--8204},
title = {{Pauli diagonal channels constant on axes}},
url = {https://iopscience.iop.org/article/10.1088/1751-8113/40/28/S22},
volume = {40},
year = {2007}
}

@article{Datta2006,
author = {Datta, N. and Fukuda, M. and Holevo, A. S.},
doi = {10.1007/s11128-006-0021-6},
issn = {1570-0755},
journal = {Quantum Info. Process. },
month = {jun},
number = {3},
pages = {179--207},
title = {{Complementarity and Additivity for Covariant Channels}},
url = {http://link.springer.com/10.1007/s11128-006-0021-6},
volume = {5},
year = {2006}
}

@article{Siudzinska2020,
author = {Siudzi{\'{n}}ska, Katarzyna},
doi = {10.1103/PhysRevA.101.062323},
issn = {2469-9926},
journal = {Phys. Rev. A},
month = {jun},
number = {6},
pages = {062323},
title = {{Geometry of generalized Pauli channels}},
url = {https://link.aps.org/doi/10.1103/PhysRevA.101.062323},
volume = {101},
year = {2020}
}

@article{Amosov2007,
author = {Amosov, Grigori G.},
doi = {10.1063/1.2406054},
issn = {0022-2488},
journal = {J. Math. Phys.},
month = {jan},
number = {1},
title = {{On Weyl channels being covariant with respect to the maximum commutative group of unitaries}},
url = {https://pubs.aip.org/jmp/article/48/1/012104/290461/On-Weyl-channels-being-covariant-with-respect-to},
volume = {48},
pages = {012104},
year = {2007}
}

@article{Li2025,
author = {Li, Yahui and Pollmann, Frank and Read, Nicholas and Sala, Pablo},
doi = {10.1103/PhysRevX.15.011068},
issn = {2160-3308},
journal = {Phys. Rev. X},
month = {mar},
number = {1},
pages = {011068},
publisher = {American Physical Society},
title = {{Highly Entangled Stationary States from Strong Symmetries}},
url = {https://link.aps.org/doi/10.1103/PhysRevX.15.011068},
volume = {15},
year = {2025}
}

@article{Siudzinska2018,
author = {Siudzi{\'{n}}ska, Katarzyna and Chru{\'{s}}ci{\'{n}}ski, Dariusz},
doi = {10.1063/1.5013604},
issn = {0022-2488},
journal = {J. Math. Phys.},
month = {mar},
number = {3},
title = {{Quantum channels irreducibly covariant with respect to the finite group generated by the Weyl operators}},
url = {http://dx.doi.org/10.1063/1.5013604},
volume = {59},
pages= {033508},
year = {2018}
}

@book{Watrous2018,
address = {Cambridge},
author = {Watrous, John},
publisher = {Cambridge University Press},
title = {{The Theory of Quantum Information}},
url= {https://www.cambridge.org/core/books/theory-of-quantum-information/AE4AA5638F808D2CFEB070C55431D897},
year = {2018}
}

@article{Poshtvan2022,
author = {Poshtvan, Abbas and Karimipour, Vahid},
doi = {10.1103/PhysRevA.106.062408},
issn = {2469-9926},
journal = {Phys. Rev. A},
month = {dec},
number = {6},
pages = {062408},
publisher = {American Physical Society},
title = {{Capacities of the covariant Pauli channel}},
url = {https://link.aps.org/doi/10.1103/PhysRevA.106.062408},
volume = {106},
year = {2022}
}

@article{Wagner2022,
author = {Wagner, Thomas and Kampermann, Hermann and Bru{\ss}, Dagmar and Kliesch, Martin},
doi = {10.22331/q-2022-09-19-809},
issn = {2521-327X},
journal = {Quantum},
month = {sep},
pages = {809},
title = {{Pauli channels can be estimated from syndrome measurements in quantum error correction}},
url = {https://quantum-journal.org/papers/q-2022-09-19-809/},
volume = {6},
year = {2022}
}

@article{UrRehman2024,
author = {{Ur Rehman}, Junaid and Al-Hraishawi, Hayder and Duong, Trung Q. and Chatzinotas, Symeon and Shin, Hyundong},
doi = {10.1109/TCOMM.2023.3341886},
issn = {15580857},
journal = {IEEE Trans. Commun.},
month = {apr},
number = {4},
pages = {2079--2089},
publisher = {IEEE},
title = {{On Estimating Time-Varying Pauli Noise}},
url = {https://ieeexplore.ieee.org/document/10354432/},
volume = {72},
year = {2024}
}

@article{Puchaa2019,
author = {Pucha{\l}a, Zbigniew and Rudnicki, {\L}ukasz and {\.{Z}}yczkowski, Karol},
doi = {10.1016/j.physleta.2019.04.057},
issn = {03759601},
journal = {Phys. Lett. A},
month = {jul},
number = {20},
pages = {2376--2381},
title = {{Pauli semigroups and unistochastic quantum channels}},
url = {https://linkinghub.elsevier.com/retrieve/pii/S0375960119303901},
volume = {383},
year = {2019}
}

@article{Cattaneo2025Dilations,
author = {Cattaneo, Marco},
doi = {10.1103/PhysRevA.111.022209},
issn = {2469-9926},
journal = {Phys. Rev. A},
month = {feb},
number = {2},
pages = {022209},
title = {{Strong symmetries in collision models and physical dilations of covariant quantum maps}},
url = {https://link.aps.org/doi/10.1103/PhysRevA.111.022209},
volume = {111},
year = {2025}
}

@article{DeLeon2022,
author = {de Leon, Jose Alfredo and Fonseca, Alejandro and Leyvraz, Fran{\c{c}}ois and Davalos, David and Pineda, Carlos},
doi = {10.1103/PhysRevA.106.042604},
issn = {2469-9926},
journal = {Phys. Rev. A},
month = {oct},
number = {4},
pages = {042604},
publisher = {American Physical Society},
title = {{Pauli component erasing quantum channels}},
url = {https://doi.org/10.1103/PhysRevA.106.042604 https://link.aps.org/doi/10.1103/PhysRevA.106.042604},
volume = {106},
year = {2022}
}

@book{nielsenchuang,
address = {Cambridge},
  title={Quantum Computation and Quantum Information: 10th Anniversary Edition},
  author={Nielsen, Michael A and Chuang, Isaac},
  year={2010},
  publisher={Cambridge University Press},
  city={Cambridge},
url ={https://www.cambridge.org/highereducation/books/quantum-computation-and-quantum-information/01E10196D0A682A6AEFFEA52D53BE9AE#contents}
}

@article{VomEnde2023,
author = {vom Ende, Frederik},
issn = {1230-1612},
journal = {Open Sys. Info. Dyn.},
month = {mar},
pages = {2350003 },
title = {{Quantum-Dynamical Semigroups and the Church of the Larger Hilbert Space}},
url = {https://www.worldscientific.com/doi/10.1142/S1230161223500038},
volume = {30},
year = {2023}
}

@book{Bengtsson2017,
address = {Cambridge},
author = {Bengtsson, Ingemar and \.{Z}yczkowski, Karol},
isbn = {9781107026254},
publisher = {Cambridge University Press},
title = {{Geometry of Quantum States - An Introduction to Quantum Entanglement, 2nd Edition}},
year = {2017},
url = {https://www.cambridge.org/core/books/geometry-of-quantum-states/4BA9DCEED5BB16B222A917EAAAD17028}
}

@article{Stinespring1955,
author = {Stinespring, W. Forrest},
doi = {10.2307/2032342},
issn = {00029939},
journal = {Proceedings of the American Mathematical Society},
month = {apr},
number = {2},
pages = {211},
title = {{Positive Functions on C$*$-Algebras}},
url = {https://www.jstor.org/stable/2032342?origin=crossref},
volume = {6},
year = {1955}
}

@book{vacchini2024open,
  title={Open Quantum Systems - Foundations and Theory},
  author={Vacchini, Bassano},
  publisher={Springer},
  year={2024},
address={Cham},
  url={https://link.springer.com/book/10.1007/978-3-031-58218-9}
}

@article{Holevo1993,
author = {Holevo, A.S.},
doi = {10.1016/0034-4877(93)90014-6},
issn = {00344877},
journal = {Rep. Math. Phys.},
month = {apr},
number = {2},
pages = {211--216},
title = {{A note on covariant dynamical semigroups}},
url = {https://linkinghub.elsevier.com/retrieve/pii/0034487793900146},
volume = {32},
year = {1993}
}

@article{Wei2018,
author = {Wei, Shi-jie and Xin, Tao and Long, Gui-lu},
issn = {1674-7348},
journal = {Sci. China Phys. Mech.},
month = {jul},
number = {7},
pages = {70311},
title = {{Efficient universal quantum channel simulation in IBM's cloud quantum computer}},
url = {http://link.springer.com/10.1007/s11433-017-9181-9},
volume = {61},
year = {2018}
}

@article{Xin2017,
author = {Xin, Tao and Wei, Shi Jie and Pedernales, Julen S. and Solano, Enrique and Long, Gui Lu},
issn = {24699934},
journal = {Phys. Rev. A},
month = {dec},
number = {6},
pages = {062303},
title = {{Quantum simulation of quantum channels in nuclear magnetic resonance}},
url = {https://link.aps.org/doi/10.1103/PhysRevA.96.062303},
volume = {96},
year = {2017}
}

@article{Cusumano2022,
author = {Cusumano, Stefano},
issn = {1099-4300},
journal = {Entropy},
month = {sep},
number = {9},
pages = {1258},
title = {{Quantum Collision Models: A Beginner Guide}},
url = {https://www.mdpi.com/1099-4300/24/9/1258},
volume = {24},
year = {2022}
}

@article{Garcia-Perez2020,
author = {Garc{\'{i}}a-P{\'{e}}rez, Guillermo and Rossi, Matteo A. C. and Maniscalco, Sabrina},
issn = {2056-6387},
journal = {npj Quantum Inf.},
month = {dec},
number = {1},
pages = {1},
publisher = {Springer US},
title = {{IBM Q Experience as a versatile experimental testbed for simulating open quantum systems}},
url = {http://dx.doi.org/10.1038/s41534-019-0235-y http://arxiv.org/abs/1906.07099 http://www.nature.com/articles/s41534-019-0235-y},
volume = {6},
year = {2020}
}

@article{Barreiro2011,
author = {Barreiro, Julio T and M{\"{u}}ller, Markus and Schindler, Philipp and Nigg, Daniel and Monz, Thomas and Chwalla, Michael and Hennrich, Markus and Roos, Christian F. and Zoller, Peter and Blatt, Rainer},
issn = {0028-0836},
journal = {Nature},
month = {feb},
number = {7335},
pages = {486--491},
title = {{An open-system quantum simulator with trapped ions}},
url = {http://www.nature.com/articles/nature09801},
volume = {470},
year = {2011}
}

@article{Han2021,
author = {Han, J and Cai, W and Hu, L and Mu, X and Ma, Y and Xu, Y and Wang, W. and Wang, H and Song, Y. P. and Zou, C.-L. and Sun, L},
issn = {0031-9007},
journal = {Phys. Rev. Lett.},
month = {jul},
number = {2},
pages = {020504},
title = {{Experimental Simulation of Open Quantum System Dynamics via Trotterization}},
url = {https://link.aps.org/doi/10.1103/PhysRevLett.127.020504},
volume = {127},
year = {2021}
}

@article{Schindler2013,
author = {Schindler, P and M{\"{u}}ller, Markus and Nigg, D. and Barreiro, J. T. and Martinez, E. A. and Hennrich, M. and Monz, T. and Diehl, S. and Zoller, P. and Blatt, R.},
issn = {1745-2473},
journal = {Nat. Phys.},
month = {jun},
number = {6},
pages = {361--367},
publisher = {Nature Publishing Group},
title = {{Quantum simulation of dynamical maps with trapped ions}},
url = {http://www.nature.com/articles/nphys2630},
volume = {9},
year = {2013}
}

@article{Faist2021,
author = {Faist, Philippe and Berta, Mario and Brandao, Fernando G. S. L.},
issn = {0010-3616},
journal = {Commun. Math. Phys.},
month = {jun},
number = {3},
pages = {1709--1750},
title = {{Thermodynamic Implementations of Quantum Processes}},
url = {http://arxiv.org/abs/1911.05563 http://dx.doi.org/10.1007/s00220-021-04107-w https://link.springer.com/10.1007/s00220-021-04107-w},
volume = {384},
year = {2021}
}

@phdthesis{MarvianMashhad2012,
author = {{Marvian Mashhad}, Iman},
school = {University of Waterloo},
title = {{Symmetry, Asymmetry and Quantum Information}},
url = {https://hdl.handle.net/10012/7088%0Ahttps://uwspace.uwaterloo.ca/handle/10012/7088},
year = {2012}
}

@article{Marvian2014,
author = {Marvian, Iman and Spekkens, Robert W.},
issn = {2041-1723},
journal = {Nat. Commun.},
month = {sep},
number = {1},
pages = {3821},
publisher = {Nature Publishing Group},
title = {{Extending Noether's theorem by quantifying the asymmetry of quantum states}},
url = {http://www.nature.com/articles/ncomms4821},
volume = {5},
year = {2014}
}

@article{Holevo1996,
author = {Holevo, A. S.},
issn = {0022-2488},
journal = {J. Math. Phys.},
month = {apr},
number = {4},
pages = {1812--1832},
title = {{Covariant quantum Markovian evolutions}},
url = {http://aip.scitation.org/doi/10.1063/1.531481},
volume = {37},
year = {1996}
}

@Inbook{Vacchini2009,
author="Vacchini, Bassano",
editor="Br{\"u}ning, Erwin
and Petruccione, Francesco",
title="Covariant Mappings for the Description of Measurement, Dissipation and Decoherence in Quantum Mechanics",
bookTitle="Theoretical Foundations of Quantum Info. Process.  and Communication: Selected Topics",
year="2010",
publisher="Springer Berlin Heidelberg",
address="Berlin, Heidelberg",
pages="39--77",
isbn="978-3-642-02871-7",
doi="10.1007/978-3-642-02871-7_2",
url="https://doi.org/10.1007/978-3-642-02871-7_2"
}

@article{Buca2012,
author = {Bu{\v{c}}a, Berislav and Prosen, Toma{\v{z}}},
issn = {1367-2630},
journal = {New J. Phys.},
month = {jul},
number = {7},
pages = {073007},
title = {{A note on symmetry reductions of the Lindblad equation: transport in constrained open spin chains}},
url = {http://stacks.iop.org/1367-2630/14/i=7/a=073007?key=crossref.ad6ff193b1621f25d8c8824080a405f0},
volume = {14},
year = {2012}
}

@article{Wallman2016,
author = {Wallman, Joel J. and Emerson, Joseph},
doi = {10.1103/PhysRevA.94.052325},
issn = {2469-9926},
journal = {Phys. Rev. A},
month = {nov},
number = {5},
pages = {052325},
publisher = {American Physical Society},
title = {{Noise tailoring for scalable quantum computation via randomized compiling}},
url = {https://link.aps.org/doi/10.1103/PhysRevA.94.052325},
volume = {94},
year = {2016}
}

@article{Lacroix2024a,
author = {Lacroix, Thibaut and Cilluffo, Dario and Huelga, Susana F. and Plenio, Martin B.},
doi = {10.1038/s42005-025-02201-2},
issn = {2399-3650},
journal = {Commun. Phys.},
month = {jul},
number = {1},
pages = {268},
title = {{Making quantum collision models exact}},
url = {https://www.nature.com/articles/s42005-025-02201-2},
volume = {8},
year = {2025}
}

@article{Flammia2020,
author = {Flammia, Steven T. and Wallman, Joel J.},
doi = {10.1145/3408039},
issn = {2643-6809},
journal = {ACM Trans. Quantum Comput.},
month = {dec},
number = {1},
pages = {1--32},
title = {{Efficient Estimation of Pauli Channels}},
url = {https://dl.acm.org/doi/10.1145/3408039},
volume = {1},
year = {2020}
}

@article{gutierrez2015comparison,
  title={Comparison of a quantum error-correction threshold for exact and approximate errors},
  author={Guti{\'e}rrez, Mauricio and Brown, Kenneth R},
  journal={Phys. Rev. A},
  volume={91},
  url = {https://journals.aps.org/pra/abstract/10.1103/PhysRevA.91.022335},
  number={2},
  pages={022335},
  year={2015},
  publisher={APS}
}

@phdthesis{Chen2025,
author = {Chen, Senrui},
pmid = {38252772},
school = {University of Chicago},
title = {{Topics in Pauli channel learning: quantum advantages, quantum noise characterization, and quantum error mitigation}},
url = {https://knowledge.uchicago.edu/record/14907?v=pdf},
year = {2025}
}

@article{Scutaru1979,
author = {Scutaru, H.},
journal = {Rep. Math. Phys.},
number = {1},
pages = {79--87},
title = {{Some remarks on covariant completely positive linear maps on $C^*$-algebras}},
url = {https://linkinghub.elsevier.com/retrieve/pii/0034487779900405},
volume = {16},
year = {1979}
}

@book{rivas2012open,
address={Berlin/Heidelberg},
  title={Open quantum systems},
  author={Rivas, Angel and Huelga, Susana F},
  volume={10},
  year={2012},
  publisher={Springer},
url = {https://link.springer.com/book/10.1007/978-3-642-23354-8}
}

@article{Cattaneo2023,
author = {Cattaneo, Marco and Rossi, Matteo A.C. and Garc{\'{i}}a-P{\'{e}}rez, Guillermo and Zambrini, Roberta and Maniscalco, Sabrina},
issn = {2691-3399},
journal = {PRX Quantum},
month = {mar},
number = {1},
pages = {010324},
title = {{Quantum Simulation of Dissipative Collective Effects on Noisy Quantum Computers}},
url = {https://link.aps.org/doi/10.1103/PRXQuantum.4.010324},
volume = {4},
year = {2023}
}

@article{Cattaneo2021,
author = {Cattaneo, Marco and {De Chiara}, Gabriele and Maniscalco, Sabrina and Zambrini, Roberta and Giorgi, Gian Luca},
issn = {0031-9007},
journal = {Phys. Rev. Lett.},
month = {apr},
number = {13},
pages = {130403},
publisher = {American Physical Society},
title = {{Collision Models Can Efficiently Simulate Any Multipartite Markovian Quantum Dynamics}},
url = {http://arxiv.org/abs/2010.13910 https://link.aps.org/doi/10.1103/PhysRevLett.126.130403},
volume = {126},
year = {2021}
}

@article{Cattaneo2022d,
author = {Cattaneo, Marco and Giorgi, Gian Luca and Zambrini, Roberta and Maniscalco, Sabrina},
issn = {1230-1612},
journal = {Open Sys. Info. Dyn.},
month = {sep},
number = {03},
pages = {2250015},
title = {{A Brief Journey through Collision Models for Multipartite Open Quantum Dynamics}},
url = {https://www.worldscientific.com/doi/10.1142/S1230161222500159},
volume = {29},
year = {2022}
}

@article{Burger2022,
author = {Burger, Andreas and Kwek, Leong Chuan and Poletti, Dario},
issn = {1099-4300},
journal = {Entropy},
month = {dec},
number = {12},
pages = {1766},
title = {{Digital Quantum Simulation of the Spin-Boson Model under Markovian Open-System Dynamics}},
url = {https://www.mdpi.com/1099-4300/24/12/1766},
volume = {24},
year = {2022}
}

@article{Erbanni2023,
author = {Erbanni, Rebecca and Xu, Xiansong and Demarie, Tommaso F. and Poletti, Dario},
issn = {2469-9926},
journal = {Phys. Rev. A},
month = {sep},
number = {3},
pages = {032619},
publisher = {American Physical Society},
title = {{Simulating quantum transport via collisional models on a digital quantum computer}},
url = {https://doi.org/10.1103/PhysRevA.108.032619 https://link.aps.org/doi/10.1103/PhysRevA.108.032619},
volume = {108},
year = {2023}
}

@article{Ciccarello2021,
author = {Ciccarello, Francesco and Lorenzo, Salvatore and Giovannetti, Vittorio and Palma, G Massimo},
issn = {03701573},
journal = {Phys. Rep.},
month = {apr},
pages = {1--70},
title = {{Quantum collision models: Open system dynamics from repeated interactions}},
url = {http://arxiv.org/abs/2106.11974 https://linkinghub.elsevier.com/retrieve/pii/S0370157322000035},
volume = {954},
year = {2022}
}

@article{Campbell2021a,
author = {Campbell, Steve and Vacchini, Bassano},
issn = {0295-5075},
journal = {Europhys. Lett.},
month = {mar},
number = {6},
pages = {60001},
title = {{Collision models in open system dynamics: A versatile tool for deeper insights?}},
url = {https://iopscience.iop.org/article/10.1209/0295-5075/133/60001},
volume = {133},
year = {2021}
}

@article{Keyl1999,
author = {Keyl, M. and Werner, R. F.},
issn = {0022-2488},
journal = {J. Math. Phys.},
month = {jul},
number = {7},
pages = {3283--3299},
title = {{Optimal cloning of pure states, testing single clones}},
url = {http://aip.scitation.org/doi/10.1063/1.532887},
volume = {40},
year = {1999}
}

@book{Paulsen2003,
author = {Paulsen, Vern},
address = {Cambridge},
isbn = {9780521816694},
publisher = {Cambridge University Press},
title = {{Completely Bounded Maps and Operator Algebras}},
year = {2003},
url = {https://www.cambridge.org/core/books/completely-bounded-maps-and-operator-algebras/47AF05B5F924ADE4FA30770B10050B76}
}

@article{DelRe2020,
author = {{Del Re}, Lorenzo and Rost, Brian and Kemper, A. F. and Freericks, J. K.},
issn = {2469-9950},
journal = {Phys. Rev. B},
month = {sep},
number = {12},
pages = {125112},
publisher = {American Physical Society},
title = {{Driven-dissipative quantum mechanics on a lattice: Simulating a fermionic reservoir on a quantum computer}},
url = {https://link.aps.org/doi/10.1103/PhysRevB.102.125112},
volume = {102},
year = {2020}
}

@article{Caruso2006,
author = {Caruso, F. and Giovannetti, V. and Holevo, A. S.},
issn = {13672630},
journal = {New J. Phys.},
url = { 	
https://doi.org/10.1088/1367-2630/8/12/310},
primaryClass = {quant-ph},
title = {{One-mode bosonic Gaussian channels: A full weak-degradability classification}},
volume = {8},
pages = {310},
year = {2006}
}

@article{David2024,
author = {David, I J and Sinayskiy, I and Petruccione, F},
issn = {2662-4400},
journal = {EPJ Quantum Technol.},
month = {dec},
number = {1},
pages = {14},
publisher = {The Author(s)},
title = {{Digital simulation of convex mixtures of Markovian and non-Markovian single qubit Pauli channels on NISQ devices}},
url = {https://epjquantumtechnology.springeropen.com/articles/10.1140/epjqt/s40507-024-00224-2},
volume = {11},
year = {2024}
}

@article{Albert2014,
author = {Albert, Victor V. and Jiang, Liang},
issn = {1050-2947},
journal = {Phys. Rev. A},
month = {feb},
number = {2},
pages = {022118},
title = {{Symmetries and conserved quantities in Lindblad master equations}},
url = {https://link.aps.org/doi/10.1103/PhysRevA.89.022118},
volume = {89},
year = {2014}
}

\end{document}